\documentclass[12pt,preprint]{aastex}

\usepackage{graphicx}
\usepackage{multicol}
\usepackage{wrapfig}
\usepackage{natbib}

\newcommand{\be}{\begin{equation}}
\newcommand{\ee}{\end{equation}}
\newcommand{\nn}{\mbox{} \nonumber \\ \mbox{} }
\newcommand{\ba}{\begin{eqnarray}}
\newcommand{\ea}{\end{eqnarray}}
\newcommand{\om}{\omega}
\newcommand{\Alfven}{ Alfv\'{e}n }

\newcommand\eg{\textit{e.g.\ }}

\newcommand\ie{\textit{i.e.\ }}

\newcommand{\maa}{\alpha \alpha} 
\newcommand{\mga}{\gamma \alpha}

\newcommand{\Bf}{{magnetic field}}
\newcommand{\Bfs}{{magnetic fields}}

\newcommand{\NS}{neutron star}
\newcommand{\NSs}{{neutron stars}}

\newcommand{\Ch}{Chandrasekhar}

\begin{document}

\title{Short Gamma-Ray Bursts  following mergers of a ONeMg with a CO white dwarf}

\author{Maxim Lyutikov\\
Department of Physics, Purdue University, 
 525 Northwestern Avenue,
West Lafayette, IN
47907-2036 and Department of Physics and McGill Space Institute, McGill University, 3600 University Street, Montreal, Quebec H3A 2T8, Canada \\
and\\
Silvia Toonen\\
Astronomical Institute Anton Pannekoek, University of Amsterdam, P.O. Box 94249, 1090 GE, Amsterdam}

\begin{abstract}
We discuss a scenario of short Gamma-Ray Bursts  (GRBs)  following a merger of a massive ONeMg white dwarf (WD) with a CO WD, and an ensuing  accretion induced collapse (AIC). An initial   system with the primary mass   $M_1 \sim 6-10 \, M_\odot$  and the secondary mass   $M_2 \sim 3-6\, M_\odot$ forms,  via two distinct evolutionary channels, a double degenerate  CO-ONeMg  WD system. For sufficiently large mass ratio $q \equiv M_2/M_1 > q_{crit} \sim 0.25$  the  ensuing gravitational wave-driven     mass transfer is unstable, whereby  the less massive CO WD  is disrupted and transfers   its mass to the primary  ONeMg WD on  a few orbital time scales. 
The merger product ignites shell CO burning, adding mass to the degenerate core; at the same time mass and angular momentum  is lost due to powerful winds.
For  an ONeMg  WD sufficiently close to the Chandrasekhar mass an electron-capture accretion induced collapse  (AIC) follows $\sim 10^4$ years afterwards.  We associate the prompt short GRB emission  with a direct collapse of an ONeMg WD to a \NS, without formation of an accretion disk. After the collapse the accretion of the unburnt part of the shell   onto the newly formed NS powers the extended emission (EE). During the collapse the \NS\ is spun to millisecond periods and produce long lasting relativistic winds that shock against the material lost during the  shell-burning stage,  and produce afterglow emission from the wind termination shock.
  \end{abstract}

\section{The problems with NS-NS scenario for short GRBs}
\label{NSNSproblems}

The typical duration of 
Short Gamma-Ray Bursts  (GRBs)  of $\sim 1$ second require \NS-like densities for a dominant energy release, $t \sim 1/\sqrt{G \rho}$. Merger of two \NSs\
is the leading model  \citep[\eg][]{2014ARA&A..52...43B}.
There are problems with the NS-NS paradigm, though, \citep[\eg][]{lyutikov_09}.
The most critical observations that challenge the dominant NS-NS merger paradigm is that    a  number of short GRBs show powerful  extended emission (EE) tails \cite[\eg][]{2006ApJ...643..266N,2010ApJ...717..411N} and flares at times as long as $10^5$ sec \citep[\eg\ GRB050724,][]{GehrelReview}. This runs contrary to many numerical simulations which show that the dominant energy release times scale is tens to hundreds of milliseconds \citep{2011PhRvD..83l4008H,2010CQGra..27k4105R,2011PhRvL.107e1102S,2015ApJ...806L..14P,2016ApJ...824L...6R}, many orders of magnitude shorter than the  duration of the extended tails and occurrence of flares.   A small amount   of material, $\leq 10^{-2}- 10^{-3} M_\odot$,   ejected during the merger of   unequal-mass \NSs \citep[\eg][]{2017PhRvD..95h3005S} is accreted on time-scales of 1-10 secs; it is hard to see how this can produce powerful extended emission tails. 

The suggestion that a (quasi)-stable \NS\ is produced as a result of a \NS\ - \NS\ merger \citep{BuciantGRB,2011MNRAS.413.2031M} requires that the maximal mass of a \NS\ is well in excess of $2 M_\odot$. There is a problem with this assumption.
Though the minimal \NS\ mass is $1.17  M_\odot$  (PSR J0453+1559) all the binary \NS\ systems have (the well-determined) total mass of $> 2.57 M_\odot$, see Table \ref{TT} (question mark for J1807-2500B  indicates that the companion might be a massive WD). 
During merger some energy will be lost to neutrino, reducing the gravitation mass by $\sim 0.3 M_\odot$ \citep[\eg][]{2009ApJ...690.1681D}. At the same time development of various shearing instabilities is expected to bring the newly formed \NS\ into solid body rotation, eliminating possibility of additional rotational support \citep[\eg][]{2012LRR....15....8F,2017PhRvD..95h3005S,2017RPPh...80i6901B,2016arXiv161203050P,2017CQGra..34h4002P}.
Thus, the formation of a stable \NS\ as a result of a merger of two \NSs\ requires that the equation of state allowed the existence of \NSs\ with masses $\geq 2.3 M_\odot$. But this is  the minimum  required mass of a stable NS. For the majority of mergers to produce a stable post-merger NS the equation of \NS\ matter should allow even higher masses,  $\geq 2.5 M_\odot$. 

It is not clear at the moment if centrifugal support or nuclear equations of
state may allow formation of \NSs\ with masses above $\geq 2.5 M_\odot$.
Below we   explore alternative  possibilities to produce extended emission. We  develop a model that may capture the advantages of producing a millisecond magnetar \citep{Usov92} as the central source of short GRBs, yet formed via different channel: electron capture (EC) in the core of a massive ONeMG WD in a binary  during the unstable mass transfer.

\begin{table}[h!]
\caption{Companion masses and the total mass in binary NSs with well-determined masses.  The total mass in all cases is $> 2.57 M_\odot$. It is highly unlikely that such systems form a stable NS upon merger.  Source: $www3.mpifr-bonn.mpg.de\//staff\//pfreire/NS\_masses.html$.}
\begin{tabular}  {llll}
DNS  & $M_1$ & $M_2$ & $M_{tot}$ \\
J0453+1559 & 1.559(5)& 1.174(4) & 2.734(4) \\
J0737-3039A/B & 1.3381(7) &
1.2489(7)	&  2.58708(16) \\
B1534+12 & 1.3330(2) & 1.3455(2) & 2.678463(4)\\
J17560-2251 & 1.341(7)	& 1.230(7) & 2.56999(6) \\
J1807-2500B (?)	&		1.3655(21)	& 1.2064(20) & 2.57190(73)\\
J1906+0746 & 1.291(11)	& 1.322(11) & 2.6134(3) \\
B1913+16 & 1.4398(2) & 	1.3886(2) & 2.828378(7) \\
B2127+11C &  1.358(10)	& 1.354(10) & 2.71279(13)
\label{TT}
\end{tabular}
\end{table}

\section{WD-WD mergers as short GRB engines} 

\subsection{Previous work: AIC, short GRBs and Type Ia SNe}

Most calculations of WD-WD mergers are aimed at explaining the Type Ia SNe, thus {\it looking} for detonation \citep[see ][for a recent review]{2014ARA&A..52..107M}.
  Less attention has been given to models that fail to detonate. As we argue below, failed SN Ia, that collapse via electron capture, may be related to the short GRBs.
\cite{2014MNRAS.438...14D} discussed  the results of the WD-WD mergers and argued  that there is large phase space available for WD-WD mergers to produce an  accretion induced collapse (AIC).  \cite{1985ApJ...297..531N} stressed the role of carbon ignition during WD mergers in order to produce a Type Ia SN. Thus, in order to avoid explosion, there should be little carbon in the system. 

The electron capture mechanism, the key physical process behind the AIC, was suggested by 
\cite{1976A&A....46..229C}.
The most often discussed observational evidence of AIC of WD is the alternative possibility of forming MSPs \citep[\eg][]{1991PhR...203....1B,2010MNRAS.402.1437H}. 
\cite{2013A&A...558A..39T} discussed a similar scenario that may lead to the formation of MSPs, including a list of systems that might have formed via AIC.
 
 Previously, several papers discussed a possibility, with a few variations, of AIC as the central engine of short GRBs \citep{1992ApJ...392L...9D,1999ApJ...527L..43V,2007ApJ...669..585D,2009arXiv0908.1127M,2001MNRAS.320L..45K,2006MNRAS.368L...1L,2007NJPh....9...17L,2013ApJ...762L..17P}. The dominant theme is the formation of a highly magnetized rapidly spinning \NS, that loses its rotational energy by a highly magnetized wind. For example,  \cite{2009MNRAS.396.1659M} discussed the case of fast initial rotation that results in a formation of  a disk. \cite{2013ApJ...762L..17P} discussed the possible radio signal from the AIC of a WD.

We stress somewhat different aspects of the AIC. First,  we discuss in more details the evolutionary scenarios that lead to AIC,  highlighting  the importance of one companion being a heavy ONeMg WD (to avoid detonation during unstable mass transfer), the formation of a shell-burning star, post-AIC shell accretion and post-accretion spin-down.

Regarding the origin of Type Ia SNe, our model is both somewhat independent of the active discussions on single versus double degenerate/mixed  origin of SNIa (in a sense that it can work in both cases), but  at the same time the model is closely related to the controversy. As a working  assumption, we assume the double degenerate super-\Ch\ CO-CO WDs  scenario for SNIa.  The  merger of  ONeMg-CO WDs  will likely proceed in a very different regime.

\subsection{Rates of WD mergers and short GRBs} 
The synthetic rate of CO+CO WD mergers with a combined mass exceeding the Chandrasekhar limit \citep[see e.g.][for overviews]{2014A&A...563A..83C, 2012NewAR..56..122W} approaches the most recent observational estimates of the SNIa rate  \citep{2017arXiv170304540M} by a factor of a few. The Galactic SNIa rate \citep[several $10^{-3}$ yr$^{-1}$][]{Cap01} is much higher then the rate of short GRBs \citep[several $10^{-6}$ yr$^{-1}$][]{2014ARA&A..52...43B,2012MNRAS.425.2668C}.  Thus, it is required that only a few percent of WD-WD mergers  exceeding the Chandrasekhar mass produce a short GRB.
Other works \citep[\eg][]{2007NJPh....9...17L} came to similar conclusion: only about $1\%$ of WD-WD mergers are needed to account for the short GRB rate. Our population synthesis calculations generally correspond to these findings as well, see Sect.\,\ref{sec:pop_syn}.

Thus, importantly, we are looking for a narrow parameter range in the pre-merger WDs masses and compositions, and possibly spins (and correspondingly narrow parameter range in the main sequence masses and separations) that may lead to the production of the short GRBs from the double white dwarfs (DWDs) mergers. Most DWD mergers do not lead to short GRBs.

\subsection{Spatial Distribution} 

Short GRBs approximately track stellar mass, not light like the long GRBs \citep{2014ARA&A..52...43B}. They also have wide distribution of off-set distances from the centers of the host galaxies
\citep{2010ApJ...722.1946B}. \cite{2014ARA&A..52...43B} noted a similarity between the distribution  of short GRBs and Type Ia SN progenitors. In the present model this is natural consequence as the two phenomena come from nearly the same progenitors (assuming a double degenerate model for Type Ia SNe). 



One of the possible problems for the present  model is a wide distribution of the off-set distances from the centers of the host galaxies for short GRBs \citep{2010ApJ...722.1946B}. It s not clear whether the NS-NS binaries fare much better, though \citep[][ studied  distribution of merging neutron stars]{2003MNRAS.342.1169V}. Typical velocities of NS binaries is tens of kilometers per second (Tauris, priv. comm.; only PSR B1913+16 has $240$ km s$^{-1}$). Presumably, higher kicks disrupt a binary.  These velocities are not too different from the velocity dispersion in the Galaxy, only slightly higher.  WD-WD binaries are expected to have smaller velocity, but this will be partially compensated by more extended location of the origin of the WD binaries  (many in halos and globular clusters) compared to the NS binaries (which originate from the galactic disk.)  We acknowledge a possible issue with the distribution of merging WD binaries with respect to the host galaxy as sources of GRBs.

\subsection{Hints from collapse physics}

The merger of WDs is a leading model to explain SN Ia \citep{2010ApJ...722L.157V,2010Natur.463...61P,2017arXiv170601898S}. To explain short GRBs detonation should be avoided - in the present model the accretor is a heavy ONeMg WDs, not a more common CO WD. Also, the donor is a degenerate star, \eg\ run of the mill CO WD with $M\sim 0.65 M_\odot$ (so that the final mass transfer proceeds in unstable regime). The reason is that 
ONeMg WD accreting from another WD  is more likely to collapse than explode \citep[see, \eg][]{2015MNRAS.453.1910S}  due to 
\begin{itemize}
\item Both  H-burning  and He-burning are usually unstable \citep[\eg][]{2005ApJ...628..395T}, resulting in the ejection of  the most of the accreted mass.  Thus, to bring a WD over the Chandrasekhar limit the donor must be an evolved star with no hydrogen or helium.
\item Most white dwarfs of moderate mass have a C/O composition. Carbon and oxygen are very prone to fusion reactions. The neutronisation thresholds for O, Ne, Mg are all lower than for C, implying that (i) they are more susceptible to collapse; (ii) are less likely to to trigger C burning/SN Ia explosion.
\end{itemize}

\subsection{Evolutionary scenario: pre-merger evolutionary tracks }
\label{sec:ev}

\begin{figure}[h!]
\centering
\includegraphics[width=.5\textwidth]{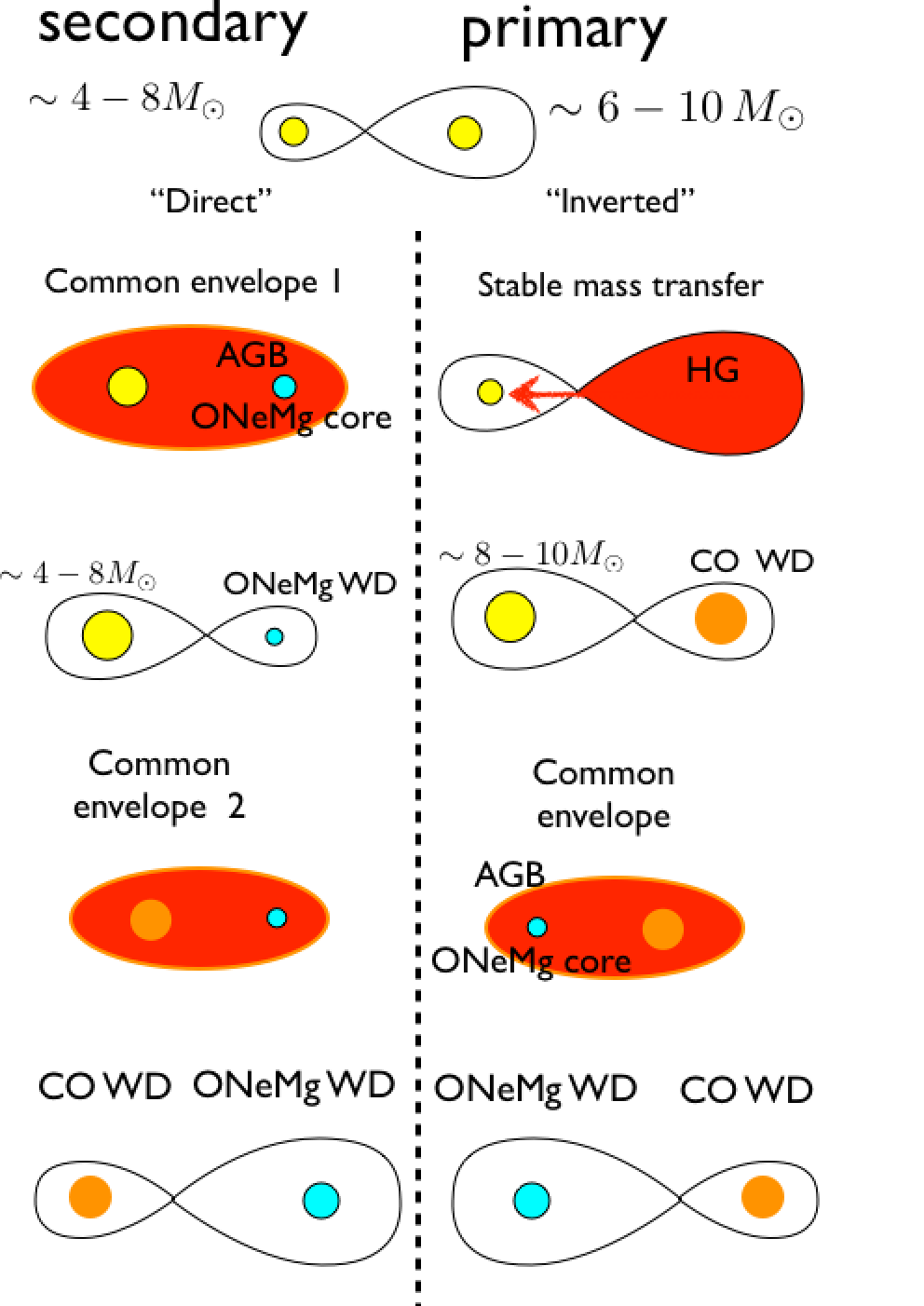}
\caption{Pre-merger  evolutionary scenarios: direct channel (left column) and inverted channel (right column).}
\label{outline-WD} 
\end{figure}

In Fig. \ref{outline-WD} we picture the evolutionary scenario leading to the short GRB.
There are two evolutionary paths that can lead to the formation of a ONeMg-CO WD binary, that we call direct and inverted. In the direct scenario, applicable to large initial separations, the ONeMg WD is formed first, while in the inverted scenario,  applicable to smaller initial separations, the primary transfers a lot of mass onto the secondary, so that the ONeMg WD is formed second. Let us discuss these scenarios in turn.

 
\subsubsection{Direct formation of  ONeMg-CO WD binary, Fig. \ref{outline-WD} left column.}

\begin{itemize}
\item {\bf  First Common Envelope phase and formation of  a ONeMg WD}. The binary starts with   two unequal mass main sequence stars ,  Fig. \ref{minit}, with the primary near the upper limit for ONeMg WD formation,  $M_1 \approx 7-10  M_\odot$ \citep{2012ARA&A..50..107L}. \citep[The primary should avoid  EC SNe, see ][]{2004ApJ...612.1044P}. 
Initial separation is  large high, on the order of hundreds/thousands  of Solar radii, Fig. \ref{ma}. As the primary starts evolving, its radius increases, up to few  $10^3 R_\odot$ - for smaller separations the system will enter the First Common Envelope stage, CE-I \citep{2013A&ARv..21...59I}.   For sufficiently large initial separation the primary will enter the CE-I at the (super-)AGB stage \citep{2006A&A...448..717S}, so its core is already a ONeMg WD. During the  CE-I the orbital separation decreases, while the mass of the primary reduces to $\sim 1.3 M_\odot$ both due to wind and loss of envelope during the  CE-I.
 The mass of the secondary does not increases  much \citep{2013A&ARv..21...59I}.

\item {\bf  Second  Common Envelope phase.} The primary is an ONeMg WD with  $\sim 1.3 M_\odot$, secondary is a Main Sequence star with $3-5 M_\odot$ The companion starts evolving, enters the giant branch  and  fills its Roche lobe. Mass loss via Roche lobe overflow results in expansion of the star,  dynamically unstable  mass transfer and the formation of a second common envelope stage, CE-II \citep{1971ARA&A...9..183P,1993PASP..105.1373I,2013A&ARv..21...59I}. The system can survive the CE-II phase \citep{2013A&ARv..21...59I}. After this stage, the system is going to consist of a ONeMg primary and CO secondary WDs.

We note that massive stars can instigate more than one phase of mass transfer. 
Once mass transfer as discussed above ceases, the old donor star evolves further. 
Helium burning takes place in its core or in a shell surrounding the core, which in its turn is surrounded by an envelope of hydrogen-poor helium-rich material. 
As this star evolves further, it can fill its Roche lobe again, leading to Case BB RLO \citep{2012MNRAS.425.1601T}. This additional post-mass-transfer phase is a result of the naked helium star (the stripped core of the original  donor star) filling its Roche lobe when it expands to become a giant during helium shell burning. For simplicity we do not include it in Fig.\,\ref{outline-WD}. It is included in the simulations described in Sect.\,\ref{sec:pop_syn}.

\end {itemize}

\subsubsection{Inverted formation of  ONeMg-CO WD binary, Fig. \ref{outline-WD} left column.} 
\begin{itemize}
\item {\bf Stable mass transfer.}  At smaller initial smaller separations,  the mass transfer can take a form of Roche Lobe overflow, at the Hertzsprung-gap stage of the primary, with a large amount of mass transferred to the secondary 
 \citep{2012A&A...546A..70T,2010A&A...515A..89M}. As a result, the primary forms a CO WD, while the secondary's mass reaches $\geq 7 M_\odot$.

\item {\bf   Common Envelope phase.} The primary is a CO WD with  $\sim 0.9 M_\odot$, secondary is a Main Sequence star with $7-10 M_\odot$. The companion starts evolving, enters the red giant branch,  fills its Roche lobe;  dynamically unstable  mass transfer leads to the formation of a second common envelope stage. Eventually, the secondary forms a ONeMg WD.
After this stage, the system consists of a ONeMg secondary   and   CO  WDs primary.

\end {itemize}

\begin{figure}[h!]
\centering
\includegraphics[width=.99\textwidth]{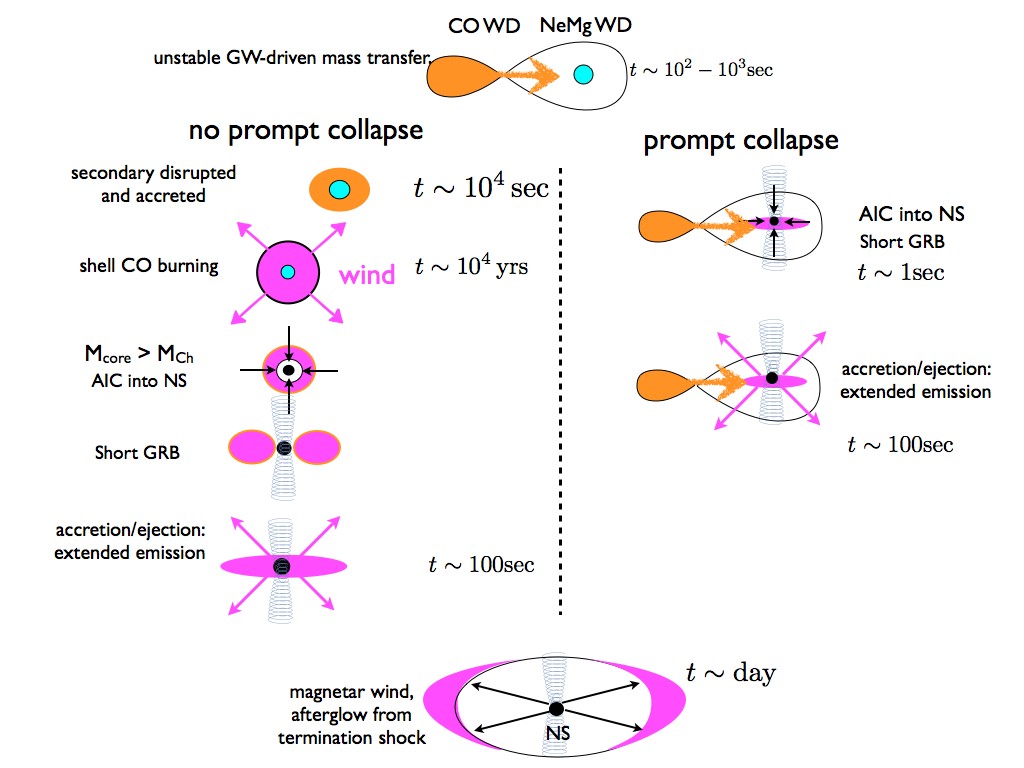}
\caption{Post-merger  evolutionary scenarios. As our basic scenario we chose left column - no prompt collapse, CO shell burning  and AIC after the core exceeds $M_{Ch}$.}
\label{outline-WD-2} 
\end{figure}

\subsection{WD-WD mergers}

Let us next discuss, somewhat independently from the previous  discussion of evolutionary scenarios, how mergers of ONeMg and CO WDs may lead to the production of short GRBs. (The case of  less massive WD being  a He WD  is also possible, but disfavor this case since  (i) accretion of He is likely to proceed explosively \citep{2013ApJ...770L...8P}; (ii) smaller mass ratios are more likely to produce stable mass transfer - see, though \cite{2012ApJ...748...35S}.) 
In the scenario we describe below many steps are controversial, and not all are calculated in detail in this work, or generally agreed upon. Yet the the following discussion is a
 reasonable description of what may happen.

\subsubsection{ONeMg-CO WD merger with shell burning, Fig. \ref{outline-WD-2}, left column}

\begin{itemize}
\item {\bf Gravitational-wave driven unstable mass transfer.} For sufficiently close WD binary, emission of gravitational waves leads to the orbital shrinking; the CO  WD starts to fill its Roche lobe, and starts transferring mass onto the primary.  As long as the mass ratio is above critical $q > q_c\approx 0.25$, the mass transfer is unstable - as matter flows from the less massive CO WD   onto the more massive ONeMg WD, the orbit expands. 
For example,  \cite{2004MNRAS.350..113M}  discuss different regimes of mass transfer in DD systems, they find that for $M_\odot \sim 1.1-1.3 M_\odot$ the mass transfer can be unstable for $q \geq 0.25$ (see also simulations  by  \cite{2012ApJ...757...76S} and discussion by \cite{2017ApJS..229...27M}).
At the same time, the CO WD expands as well (less massive WDs have larger radii). In case of unstable mass transfer the expansion of the WD over-compensates for the orbital expansion. As a result, the companion is disrupted on few orbital time scales \citep[\eg][]{2006ApJ...643..381D}.

 Most  interacting double WD binaries are likely to merge, as oppose to experience dynamically stable mass transfer \citep{2015ApJ...805L...6S}. We assume that the mass transferred onto the primary ONeMg WD is not detonated  in SN Ia event \citep[ONeMg is hard to ignite; \eg][]{2012ApJ...746...62R}, nor  that the shell is ignited in nova-like outburst. \footnote{Note that even if mass transfer between a He WD and a ONe WD is unstable, the contribution to the merger rate is a few percent at best, as only few He-ONe DWDs are formed.}
Thus, during the   unstable mass transfer  the companion is fully disrupted and accreted onto the collapsed primary 
\citep{1987fbs..conf..445W,1988ApJ...324..355I}.

\item {\bf Shell burning above ONeMg core.} After the disrupted CO WD is accreted, the shell is viscously spread-out over the core on a timescale of $\sim 10^4$ seconds
\cite{2007MNRAS.380..933Y,2012ApJ...748...35S}. Stable nuclear burning is ignited at the base of the shell, adding mass to the degenerate core. The ONeMg core is not ignited. At the same time powerful winds lead to mass loss and angular momentum loss from the envelope. The overall luminosity is of the order of Eddington luminosity.
At this stage there a competition between the wind mass  loss from the shell, and the addition of degenerate material to the core. Under certain condition the wind mass loss dominates, and the core never reaches $M_{Ch}$. (Magnetic White Dwarf EUVE J0317-85.5  with $ M=1.35M_\odot$ could be an example of such "near the cliff" WD that nearly underwent AIC - if few hundredths of $M_\odot$ woulds have been added. 

\item {\bf AIC.}   After the total mass of the core exceeds the Chandrasekhar limit, AIC of a WD to NS follows \citep{1985A&A...150L..21S,1987ApJ...315..229K,1990A&A...236..378M,1980PASJ...32..303M,2001MNRAS.320L..45K,2016MNRAS.458.3613S}.  The AIC proceeds inside-out - the core bounce is the first observed effect.  
Core-bounce leads to an outgoing shock that may create a weak  supernova-like explosion \citep{1992ApJ...391..228W, 2006ApJ...644.1063D,2006A&A...450..345K,2010PhRvD..81d4012A,2012ApJ...747...88N,Fryer99}. The outgoing material also has to plow through the still  remaining shell, remains of  the disrupted companion. This material will modify the observed properties of the explosion.


\item {\bf Magnetic fields and jets - the short GRB}.
The primary ONeMg WD is slowly rotating before the onset of the unstable mass transfer. It is spun up during the unstable   mass transfer, but a lot of angular momentum is lost \citep{2012ApJ...748...35S}.
  For sufficiently slow rotation before the AIC,   the collapse is direct, without formation of the accretion disk. In this case the collapse of the core proceeds  on dynamic, not viscous time scale. During the core collapse differential rotation amplifies B-field \citep{1995ASPC...72..301T} and produces an MHD jet \citep[][ give estimates of \Bf\ amplification during collapse]{2007ApJ...664..416B}.  
  On the other hand, the shell,  the disrupted  CO companion with $M \sim 0.1 - 0.5 M_\odot$ remaining after shell burning and wind loss, forms a disk that helps confine  mildly collimated  outflow.
 Since there is little envelope to collimate the outflow (only a fraction of the Solar mass, not few solar masses as in the case of core-collapse SNe), the jet is wide and 
 terminates quickly. 
These are the Short GRBs. Importantly, the expected duration is not a free-fall time for WD to collapse, but the bounce time, corresponding to the size of the proto-neutron star.

\item {\bf Continuing  accretion - the extended emission}. 
The disk formed from the shell accretes on time scale $\sim 100-1000$ seconds,
but the  primary is now a \NS. Thus, the newly formed
\NS\ experiences mild accretion rates on the viscous  time scale of a few orbital periods. This powers the EE.

\item {\bf  Wind from isolated NS - early afterglow.}  AIC may spin-up the resulting \NS\ to millisecond periods and amplify \Bf\ to magnetar values, creating conditions favorable for magnetar-driven out flows \citep[\eg][]{2008MNRAS.385.1455M}. After the disk is accreted/ejected, the nature of the collimation changes - isolated \NSs\ form equatorially, not axially collimated outflows, with power $\propto \sin^ 2 \theta$ \citep[$\theta$ is the polar angle]{1973ApJ...180L.133M}. (This important point was not stressed by previous models invoking millisecond magnetar as a central source of short GRBs. Conventionally, models of prompt emission, both in long and short GRBs, rely on axially collimated jets \citep[\eg][]{1999ApJ...524..262M,2011ApJ...732L...6R}. Intrinsically, rotating \NSs\ produce more power in the equatorial direction. If a heavy medium is present, \eg\ outer layers of a collapsing star, the outflow can be collimated \citep{2011MNRAS.413.2031M,2009MNRAS.394.1182K} in a way similar to the outflows in the Crab Nebula \citep{komissarov_03,komissarov_04}.)

As the highly relativistic wind from the newly formed NS interacts with the expelled shell, and with the pre-collapse wind, a termination shock is formed. The shock is relativistic and highly magnetized. The wind termination shock produces afterglows in a manner similar to the case of long GRBs, as discussed by \cite{2017ApJ...835..206L}.

\end{itemize}

 \subsubsection{ONeMg-CO WD merger  with prompt AIC,  Fig. \ref{outline-WD-2}, right column}
 \label{promptAIC}
 
A somewhat alternative possibility is that the  ONeMg is very close to the \Ch\ limit at the beginning of the unstable mass transfer. As the  accreted material is heated, it loses degeneracy and thus exerts little pressure on the ONeMg core. Still, if the mass of the    ONeMg WD is sufficiently close to the  \Ch\ limit this extra force may be sufficient to induce AIC before the disruption of the CO WD is completed.  The ensuing evolution will resemble NS-WD disruption events \citep[\eg][]{2017MNRAS.467.3556B}.

\section{Population synthesis}
\label{sec:pop_syn}
Using the binary population synthesis (BPS) code \texttt{SeBa} \citep{SPZ96, 2012A&A...546A..70T, Too13}, we simulate the evolution of a large number of binaries following in detail those that lead to the merger of an ONeMg and CO WD. Processes such as wind mass loss, stable \& unstable mass transfer, accretion, angular momentum loss, and gravitational wave emission are taken into account. It was shown by \cite{Too14} that the main source of uncertainty in the BPS outcomes come from the uncertainty in the input assumptions, in particular the CE-phase. For this reason, we follow \cite{2012A&A...546A..70T}, in performing two sets of population synthesis calculations using their model $\maa$ and $\mga$. For full details on the models, see \cite{2012A&A...546A..70T}. 
In short, these models differ from one another with respect to the modeling of the CE-phase. Despite the importance of this phase for the formation of compact binaries and the enormous effort of the community, the CE-phase is still poorly constrained \citep[see][for a review]{2013A&ARv..21...59I}. Commonly the CE-phase is modeled in BPS codes by energy conservation \citep{1984ApJ...277..355W}, with a parameter $\alpha$ that describes the efficiency with which orbital energy is consumed to unbind the CE. This recipe is used in model $\maa$ for every CE-phase.
An alternative model has been proposed by \citep{Nel00} in order to reproduce the observed population of double white dwarfs. This model is based on a balance of angular momentum with an efficiency parameter $\gamma$. In our model $\mga$, the $\gamma$-recipe is used unless the binary contains a compact object
or the CE is triggered by a tidal instability (rather than dynamically
unstable Roche lobe overflow, as proposed by \cite{Nel00} for detached double white dwarfs.

\begin{figure}[h!]
\centering
\includegraphics[width=.49\textwidth]{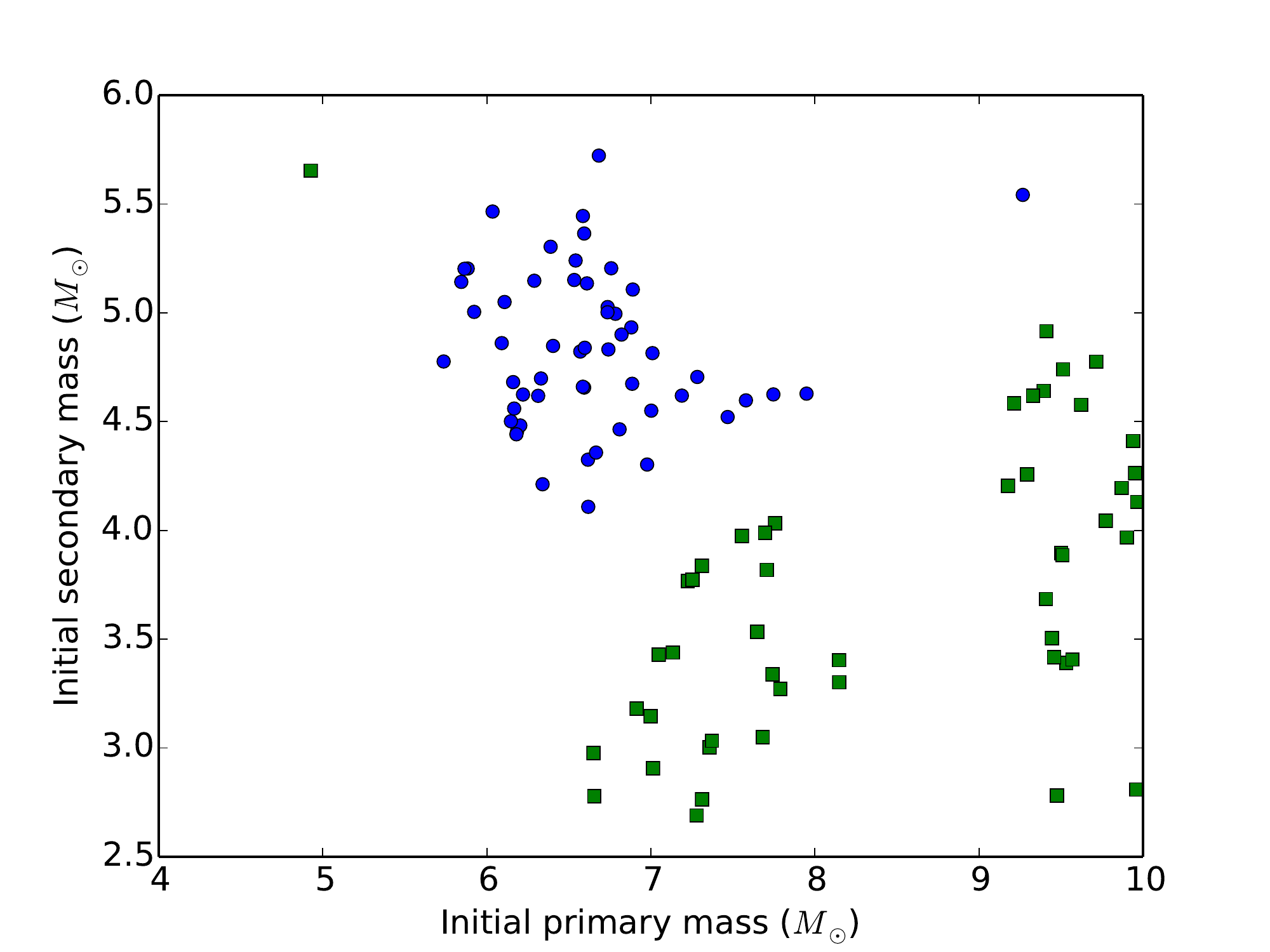}
\includegraphics[width=.49\textwidth]{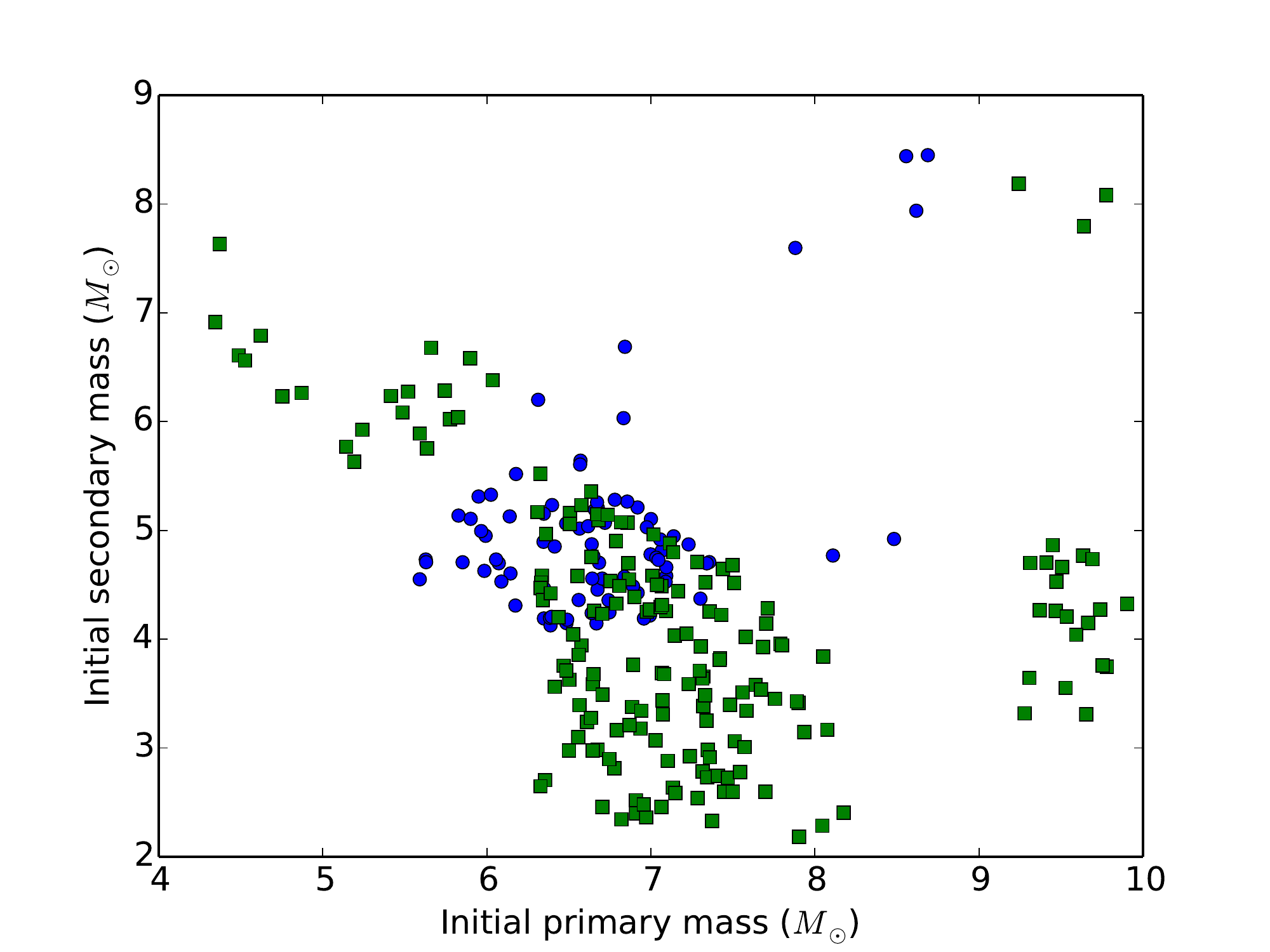}

\caption{Distribution of initial masses for model $\mga$ (left panel) and model $\maa$ (right panel). The primary represents the first formed WD, secondary the last formed WD. With green squares the systems where the ONeMg WD is formed first, with blue circles where this is the last formed WD. In all models the systems marked in blue come from tight orbits where the first phase of mass transfer is likely stable mass transfer. The systems marked in green mostly originate from wider orbits, such that that first phase of mass transfer is likely a common-envelope phase.  }
\label{minit} 
\end{figure}

\begin{figure}[h!]
\centering
\includegraphics[width=.49\textwidth]{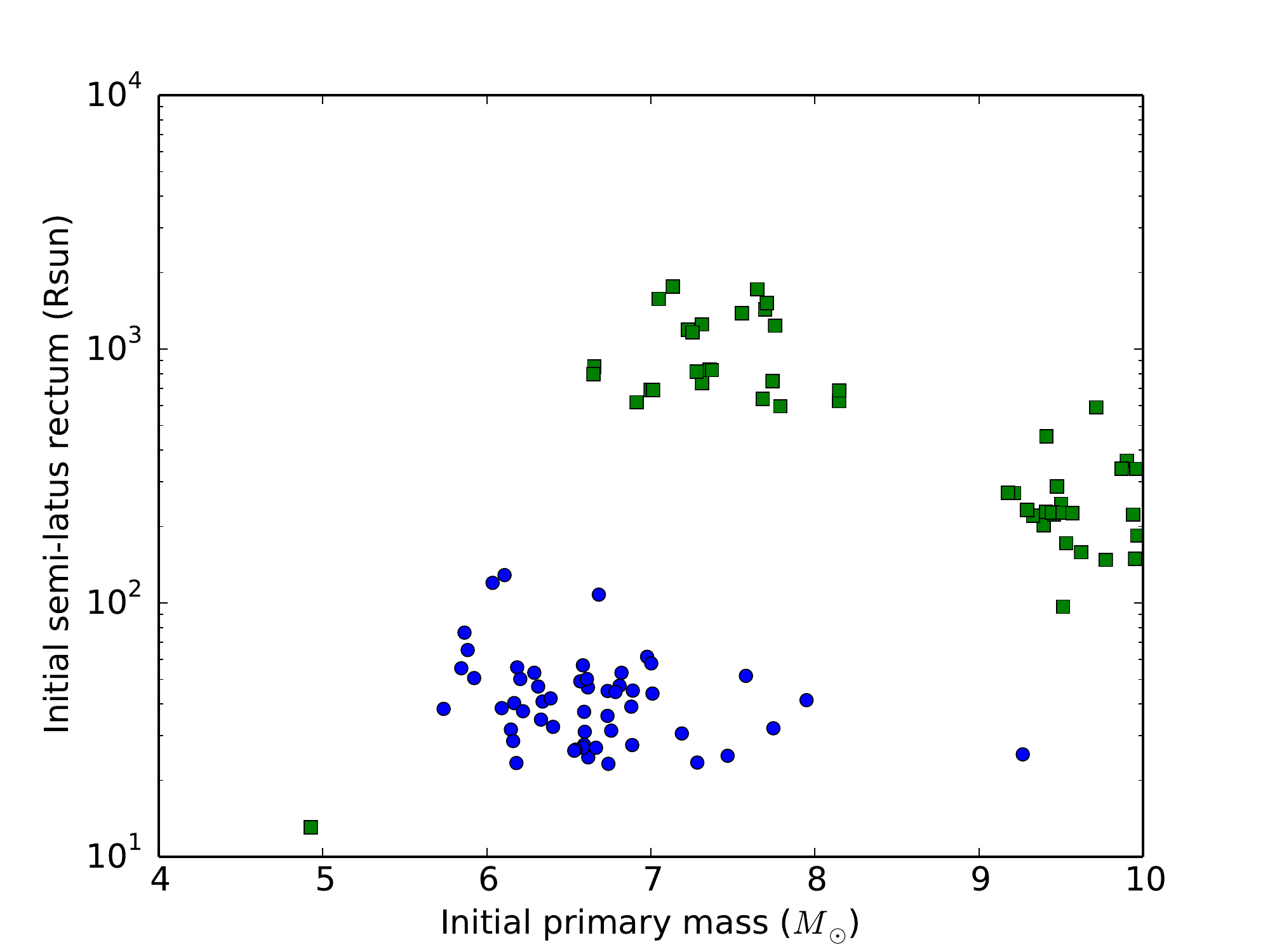}
\includegraphics[width=.49\textwidth]{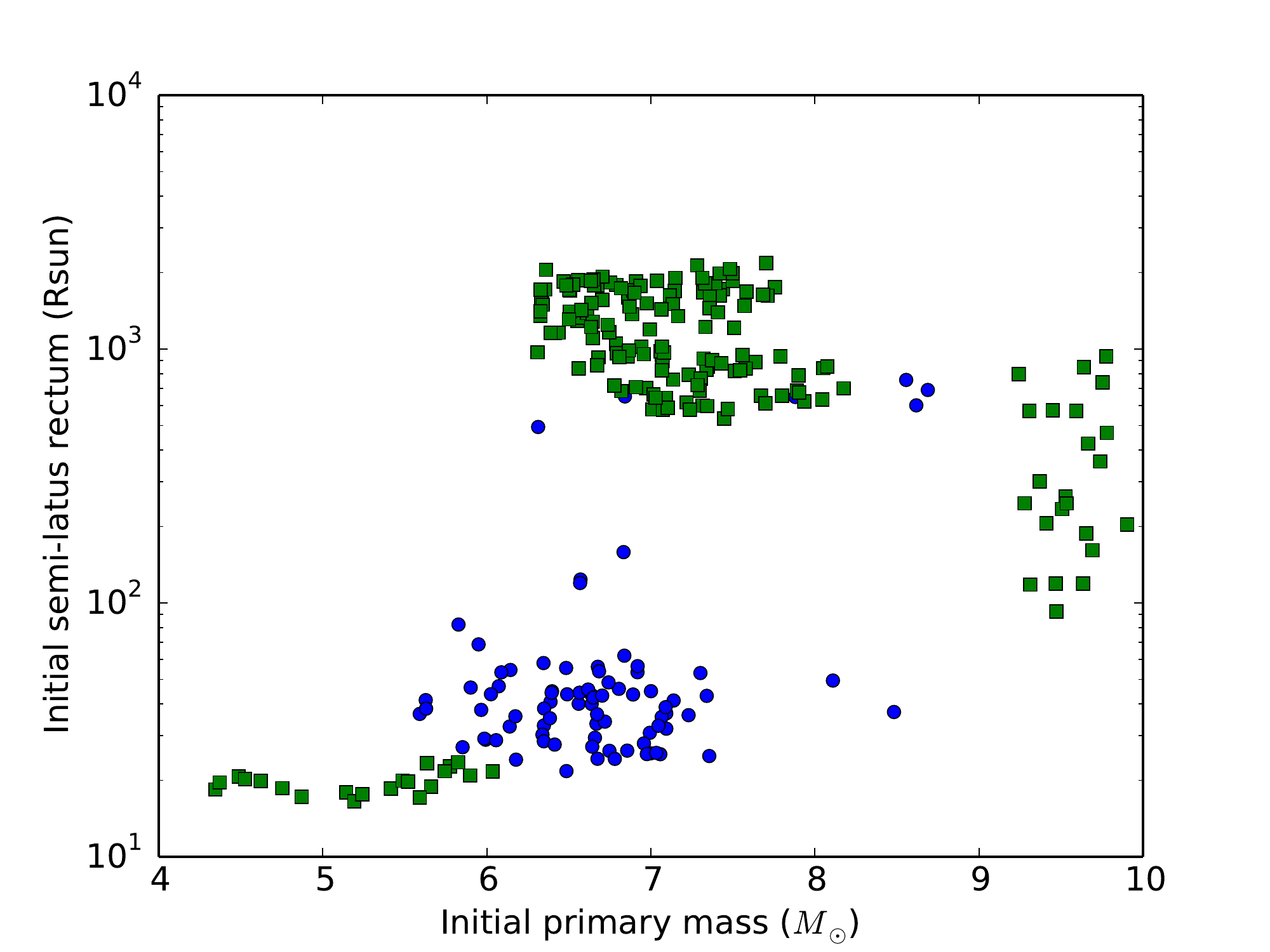}

\caption{Distribution of initial semi-latus rectum for model $\mga$ (left panel) and model $\maa$ (right panel). The color coding is the same as in Fig.\,\ref{minit}. In the case of $\maa$,  green dots at low orbital separations correspond to the systems in which the primary starts with $\sim  6 M_\odot$ and relatively small separation, so that  the first phase of mass transfer is stable. As the secondary accretes, it becomes more massive, its evolution speeds up, and it becomes a ONeMg WD while the primary is still a stripped (hydrogen poor helium-rich) nuclear burning star, which eventually becomes a WD. This is similar to the third evolution channel in \cite{2012A&A...546A..70T}.
 }
\label{ma} 
\end{figure}

Figures\,\ref{minit}-\ref{ma} show the initial parameters of binaries leading to mergers between ONeMg and CO WDs in our simulations. Every point represents a single system in the BPS simulations. The figures show that there are different evolutionary paths that can lead to an ONe-CO WD merger, however the dominant channels involve initially compact systems (in blue circles) i.e. the 'inverted channel' and initially wide systems (in green squares) i.e the 'direct' channel. For single stars, the initial mass  of the progenitor of an ONeMg WD ranges between approximately 6.5-8M$_{\odot}$ according to \texttt{SeBa}. This is similar to the range of initial masses in the direct channel where the primary forms the ONeMg WD (majority of green points in Fig.\,\ref{minit}). The progenitors of ONeMg WDs in the 'inverted' channel,  {\ie}the secondaries denoted in blue, have lower masses as these stars accrete a significant amount of mass from their companion stars.

\begin{figure}[h!]
\centering
\includegraphics[width=.49\textwidth]{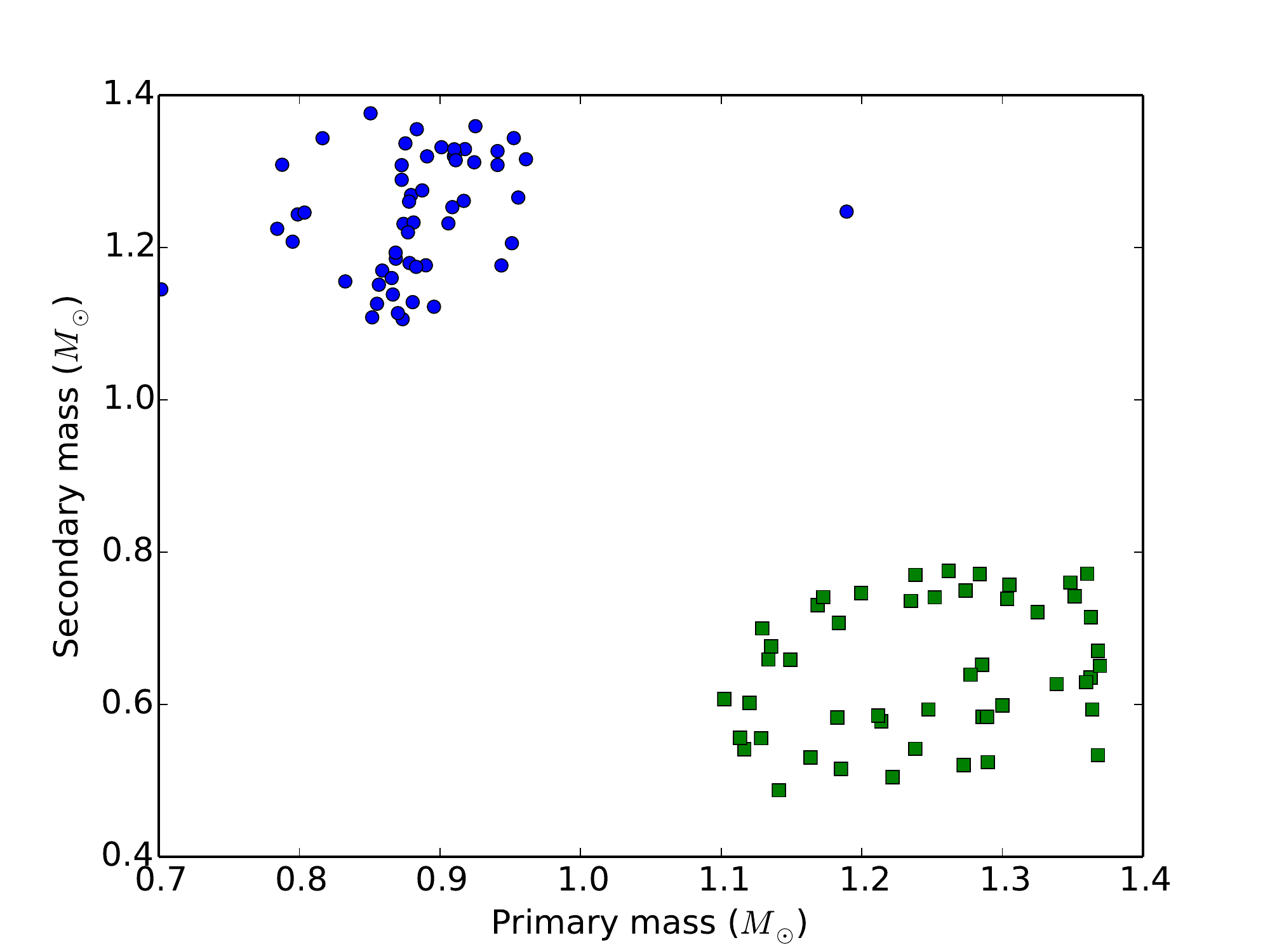}
\includegraphics[width=.49\textwidth]{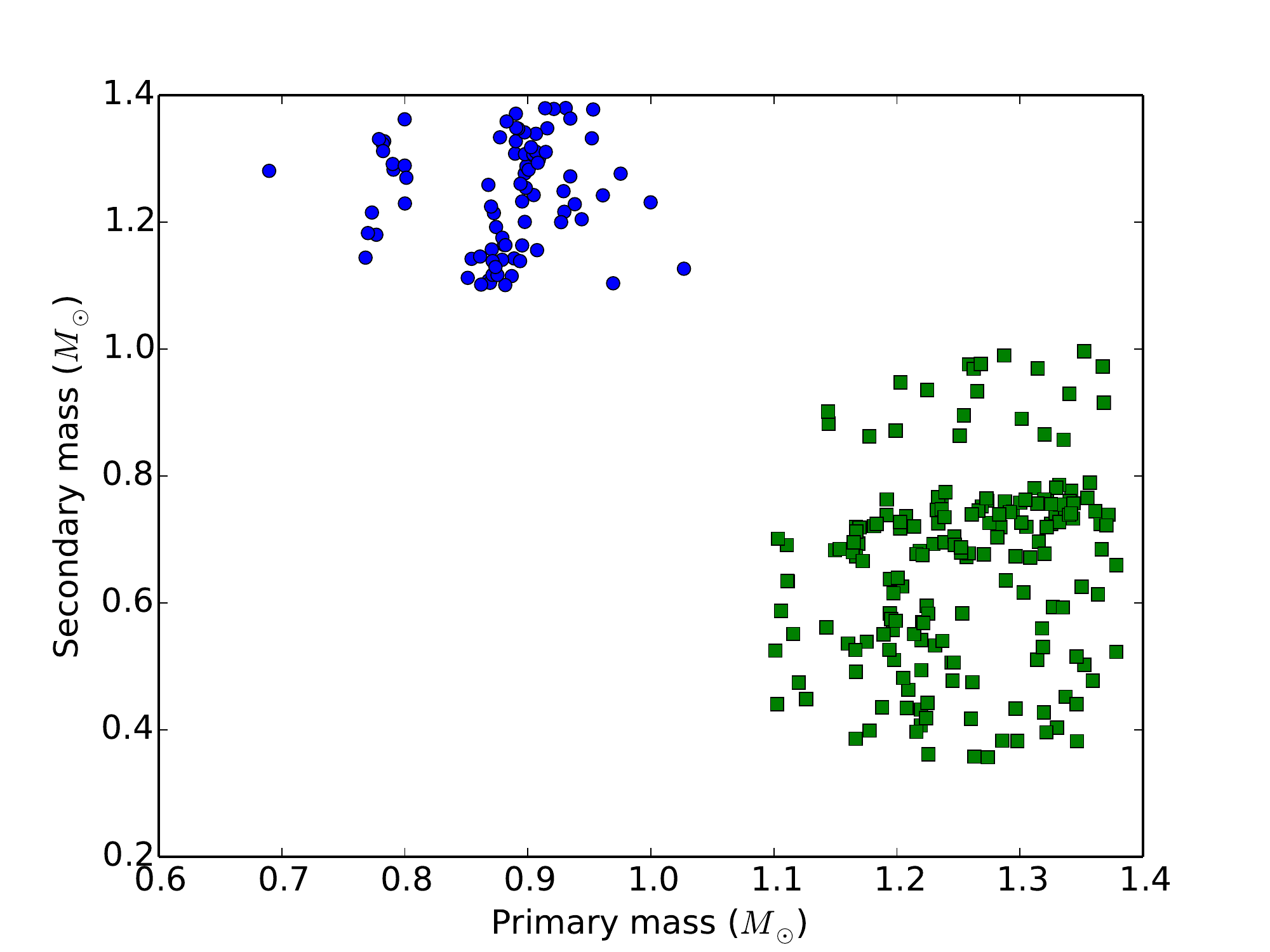}
\caption{Distribution of masses for model $\mga$ (left panel) and model $\maa$ (right panel).  The color coding is the same as in Fig.\,\ref{minit}.  }
\label{mwd} 
\end{figure}

In Fig.\,\ref{mwd}  we show the final masses of the ONeMg and CO WD that merge according to model $\maa$~and~$\mga$ respectively. The masses of the ONeMg WDs are in the range  $1.1-1.4M_{\odot}$, while the majority of CO WDs have masses in the range 0.5-0.8M$_{\odot}$. As described in Sect.\,\ref{sec:ev}, it is possible that the ONeMg WD forms before the other WD in the system (channel 'direct'), or it forms afterwards (channel 'inverted'). In model $\mga$, 48\% of merging ONe-CO DWDs go through the 'direct' channel, whereas for model $\maa$ the fraction goes up to 69\%. The masses of the CO WDs in the 'inverted' channel are systematically higher than those of the 'direct' channel.

\begin{figure}[h!]
\centering
\includegraphics[width=.49\textwidth]{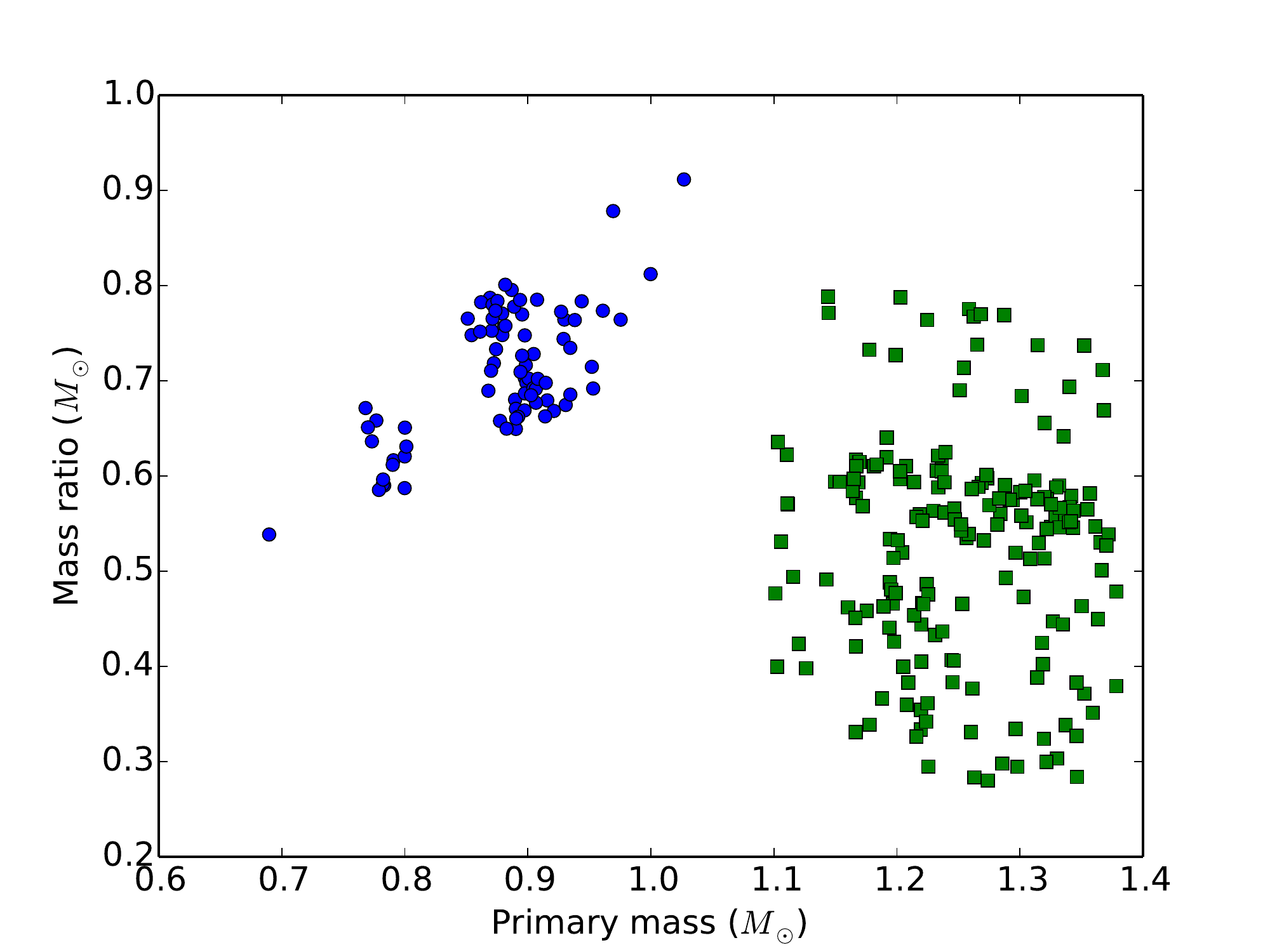}
\includegraphics[width=.49\textwidth]{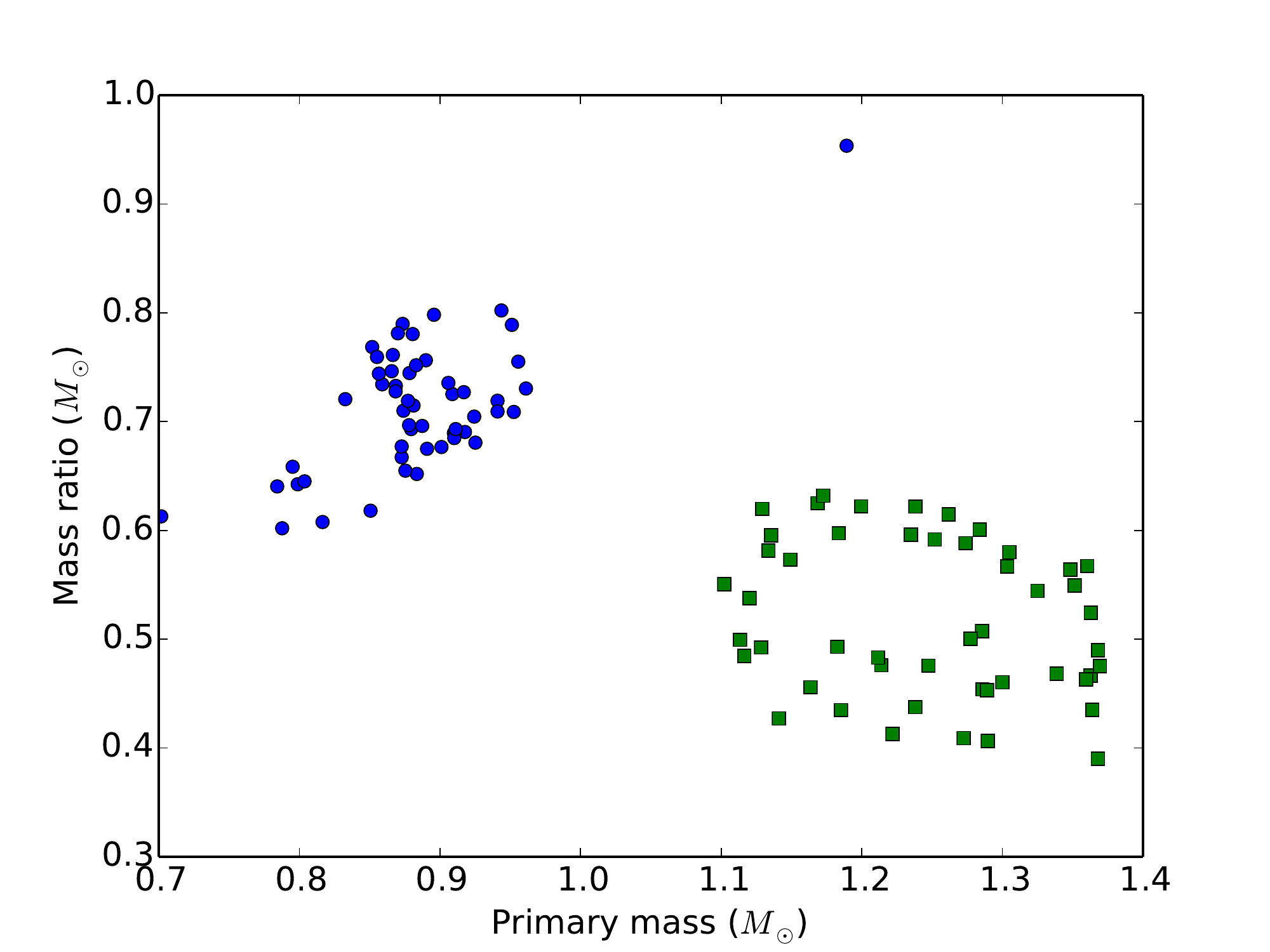}
\caption{Distribution of the mass ratio as a function of the primary mass  for model $\mga$ (left panel) and model $\maa$ (right panel).  The color coding is the same as in Fig.\,\ref{minit}.  }
\label{mwc} 
\end{figure}

In Fig.\,\ref{mwc}  we show the distribution of the mass ratio as a function of the primary mass.  For donor masses in the range $1.1-1.3 M_\odot$  \cite{2004MNRAS.350..113M} (see their Fig. 1) find that mass transfer is always unstable if the companion mass is above  $\sim 0.6$. It is always stable for $\leq 0.2-0.4$. The blue systems are well above the limit for unstable mass transfer. The green systems occupy a larger part of parameter space. The far majority of the systems have a mass ratio that make stable mass transfer unlikely. Also note that given the 'optimistic' stability limits of \cite{2004MNRAS.350..113M} the AM CVn rate is overestimated by orders of magnitude, indicating that mass transfer is likely less stable than the 'optimistic scenario'. 
In addition, the  results from \cite{2004MNRAS.350..113M}  do not take into account the effect of novae outbursts on the evolution of the systems. As shown by  \cite{2012ApJ...748...35S} these outburst have a destabilizing effect on the mass transfer.

Assuming a constant Galactic  star formation rate of $6 M_{\odot}$ yr$^{-1}$ for the last 10 Gyr\footnote{The assumed star formation history is normalized, such that the total stellar
mass corresponds to the Galactic stellar mass of $6\times 10^{10} M _{\odot}$yr$^{-1}$ \citep{Too17}}, the current merger rate of CO-ONeMg WDs is $1.9 \times 10^{-4}$yr $^{-1}$ for model $\mga$ and $5.0 \times 10^{-4}$yr $^{-1}$  for model $\maa$. This is still approximately an order of magnitude higher than the estimated short GRB rate. Thus, even among the ONeMg-CO WD mergers, only about $\sim 10\%$ need to produce a short GRB. (This conclusion strongly depends on the assumed beaming of short GRBs.)


\section{Unstable mass transfer in WD-WD binary}

\subsection{Stellar and orbital parameters}
 
Consider a WD binary with primary mass $M_{1}=1.3 M_\odot$ and companion $M_{2} = 0.65 M_\odot$ (so, $q=1/2$) and assume that the critical value for unstable mass transfer is  $q_{crit} \leq q$. As long as  $q_{crit} \leq q$ exact value of $q_{crit} $ is not important \citep[$q_{crit}$ can be as small as $0.25$, \eg][]{1990ApJ...348..647B,2006ApJ...643..381D,2007ApJ...670.1314M}; also,  the unstable mass transfer is typically not stabilized \citep{2012ApJ...748...35S}. The secondary enters the Roche lobe when the size of the WD $R_{WD}$ becomes of the order of the Roche lobe $ R_{Rl}$.
 \ba &&
R_{WD}=  \frac{(9 \pi)^{2/3}}{8} \frac{\hbar^2}{ G m_e M_{2}^{1/3}  m_p^{5/3}}
\label{RWD}
\\ && R_{Rl}=  0.46224 a_0 \left( \frac{q}{1 + q}\right)^{1/3}
\ea
 (where for qualitative estimates we use a simple expression for  the WD radius $R_{WD}$ and the size of the Roche lobe $R_{Rl}$; for more precise values see \citep{1983ApJ...268..368E}, and assume a simple ideal non-relativistic  EoS for the companion WD).
For a $M_1 = 1.3 M_\odot$ primary and $q=M_2/M_1 =1/2$ this occurs when the orbital separation $a_0$ is 
\be
a_0=\frac{1.7 {(1+q)^{1/3}} \hbar ^2}{G {M_1} ^{1/3}q^{2/3}  m_e m_p^{5/3}}
= 1.2 \times 10^9\, {\rm cm}
   \label{a0}
\ee
The gravitational wave-driven 
inspiral timescale  at that moment,
\be
\tau_G= \frac{5 a_0^4 c^5}{32 G^3 M_{1} M_{2} \left(M_{1}+M_{2}\right)}=
 2.0 \times 10^9 {\rm sec},
   \ee
   is much longer than the orbital period
   \be
   P_{orb} = \frac{2 \pi  a_0^{3/2}}{\sqrt{G \left(M_{1}+M_{2}\right)}} =16 \, {\rm sec}
   \label{Porb}
  \ee
 The estimates above use highly idealized equation of state and a simple prescription for the size of the Roche lobes;  they should be taken only as order-of-magnitude estimates.

  \subsection{ Spin evolution of the merger product}
  
  During unstable mass transfer the lighter WD is disrupted and  forms a  disk around the primary. Disk accretion at high rates creates a spreading layer - a belt-like structure on the  surface of the primary \cite{1999AstL...25..269I,2009ApJ...702.1536B,2010AstL...36..848I,2013ApJ...770...67B,2016ApJ...817...62P}. After the spreading is complete \citep[on viscose time scale of $\sim 10^4$ seconds, \eg][]{2012ApJ...748...35S}  the resulting star of $\sim 2 M_\odot$ consists of a slowly rotating degenerate ONeMg core, and a fast rotating, with period in the hundreds of seconds, non-degenerate  envelope. After the removal of the degeneracy the envelope expands  to $R_{shell} \sim$ few $10^9$ cm. The star will emit near Eddington limit and drive powerful winds.
 Angular momentum contained in the shell will be both lost to the wind, and transported to the core through the (turbulent) boundary layer. 
 
  The moment of inertial of the merger product is a sum of the moment of inertia of the core $ \eta_0  M_{1}  R_{1} ^2$ ($\eta_0$ is the gyro-ratio -  the moment of inertia divided by  $M R^2$)  and of the envelope, 
 $\approx  M_{2} R_{shell}^2$. The contribution from the envelope - the disrupted companion - dominates:
 \be
 \frac{ \eta M_1 R_1^2}{ M_2 R_{shell}^2} \approx 10^{-3} 
 \ee
 for $R_1 = 3000$ km and $q=1/2$. 
Thus, we expect that the core quickly comes into solid body rotation with the envelop.

  To estimate the angular velocity of the shell we note that after accretion the shell  quickly expands to $R_{shell}$; this expansion decreases its angular velocity. 
 As an estimate of the shell's spin we can equate the proper angular momentum of the companion, $  {\sqrt{a}  \sqrt{G }M_1^{3/2}} q/ \sqrt{1+q} $, at the point of merger  (\ref{a0})
   to the angular momentum of the shell,  $R_{shell}^2 \Omega_{shell}$. We find
 \be
   \Omega_{shell}= \frac{3^{5/6} {\pi }^{1/3}}{{2}^{7/6} \sqrt{5}}
\frac{{M_1}^{1/3} \hbar }{{q}^{1/6} \sqrt{q+1} {R_{shell}}^2 \sqrt{m_e} m_p^{5/6}} =8 \times 10^{-3} \left(\frac{R_{shell}}{5 \times 10^{9} \rm cm} \right)^{-2}
   {\rm s}^{-1},
   \label{om}
   \ee
   so that the shell rotates with a period of about  $700$ seconds. 
   This estimate agrees with more detailed calculations   \citep{1990ApJ...348..647B,2012ApJ...746...62R,2012ApJ...748...35S}. For example, \cite{2012ApJ...748...35S}  found that  as the  angular momentum redistributes, the  shell reaches nearly solid-body rotation with a period $\sim $ hundreds of seconds  \citep[Figs. 3 and 4 of][]{2012ApJ...748...35S}.

After the merger, carbon is stably burning at the base of the shell, adding degenerate material to the core. At the same time, mass is lost due to the wind. If/when 
 the mass of the core exceeds the \Ch\ mass AIC occurs. 
  We associate the prompt short GRB with an AIC {\it  directly} into the \NS, without formation of an accretion disk. This requirement comes  from the observer duration of only $\sim 1$ second for the short GBRs. If the core is nearly critically rotating before the collapse, with periods of the order of $\sim 10 $ seconds, the accretion time scale is expected to be even  longer.
  
   The requirement of a direct collapse demands   that the spin of the  ONeMg core before the AIC be not too high.  
  To find the conditions for direct collapse consider the rotation of a ONeMg core before the AIC with highly sub-Keplerian velocity on the surface,
 \be
 {\Omega_{1} R_{1} \ll  \sqrt{ G M_{1}  /R_{1}}} 
 \ee
 For the direct collapse the final \NS| should rotate with period $P_{NS} \geq 1$ msec,
For  a \NS\ with radius $R_{NS}=10$ km  the initial period of WD should be  
\be
P_{1} > (R_{1}/R_{NS})^2 P_{NS} >  90 sec.
\label{Pcrit} 
\ee
for $R_{1}/R_{NS} \approx 300$.
As we discussed above, Eq. (\ref{om}), 
 the condition (\ref{Pcrit}) is indeed satisfied following the merger - AIC of the ONeMg core is direct, without formation of the disk. (But the ensuing accretion of the envelope will lead to the formation of the disk, see \S \ref{EE}.)

\section{Collapse dynamics and \Bf\ amplification}

Next we consider the dynamics of the inside-out collapse of a polytropic sphere. First we neglect the effects of  rotation, assuming a direct radial collapse. Later, we estimate the rotationally-induced  shear and the  \Bf\ amplification within the collapsing star.


\subsection{Inside-out collapse of a polytropic sphere}

The collapse of a gaseous   sphere supported against gravity by the pressure gradients will start from the the center, launching a rarefaction wave propagating to larger radii with the local speed of sound. As a rarefaction wave reaches a given point,  the pressure support against gravity is lost, so that a given fluid element start a nearly free-fall motion in a potential generated by all the material inside this radius.  In this Section we discuss the corresponding dynamics for the case of massive WDs.

For analytical estimates, we approximate the WD as a polytropic sphere.
Starting  with Lane-Emden eq. \citep{1967aits.book.....C}
\be
{1\over x^2} \partial_x ( x^2  \partial_x \theta) = - \theta^n
\ee
where $x$ is radius, $\theta$ is density  (in dimensionless unites, whereby
$p=\kappa \rho^{1+1/n}$, $\rho= \rho_c \theta^n$, $c_s^2 = \theta c_{s,0}^2$, $c_{s,0}^2= K \rho_c^{1/n}$ and spatial scales are normalized by
\be
r_0 = \sqrt{ {n+1 \over 4 \pi G} \kappa \rho_c^{(1-n)/n} } = R_{WD}/x_{max}
\label{r0}
\ee
Time is measured in terms of 
\be
\tau = \sqrt{ n+1 \over 4 \pi G \rho_c }= 0.56 {1\over \sqrt{G  \rho_c}}
\label{tau}
\ee

\begin{figure}[h!]
\centering
\includegraphics[width=.32\textwidth]{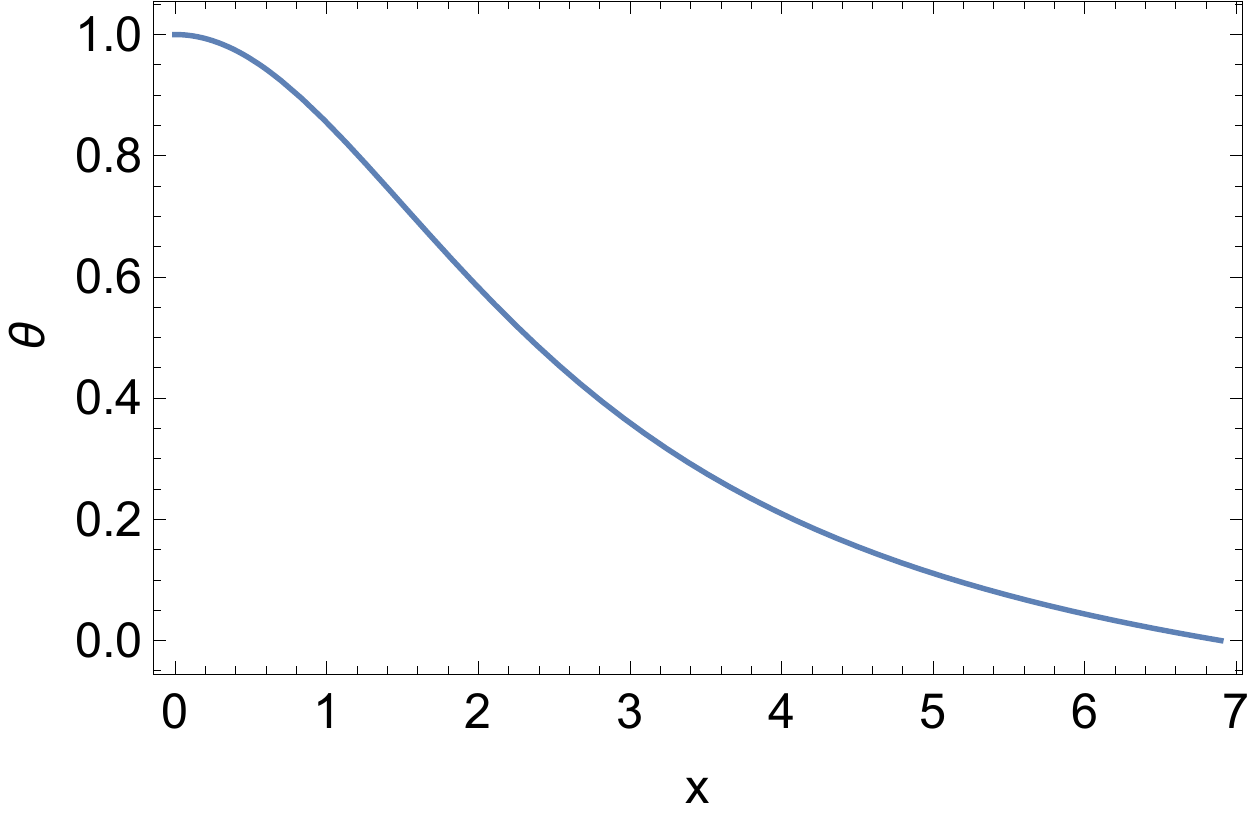}
\includegraphics[width=.32\textwidth]{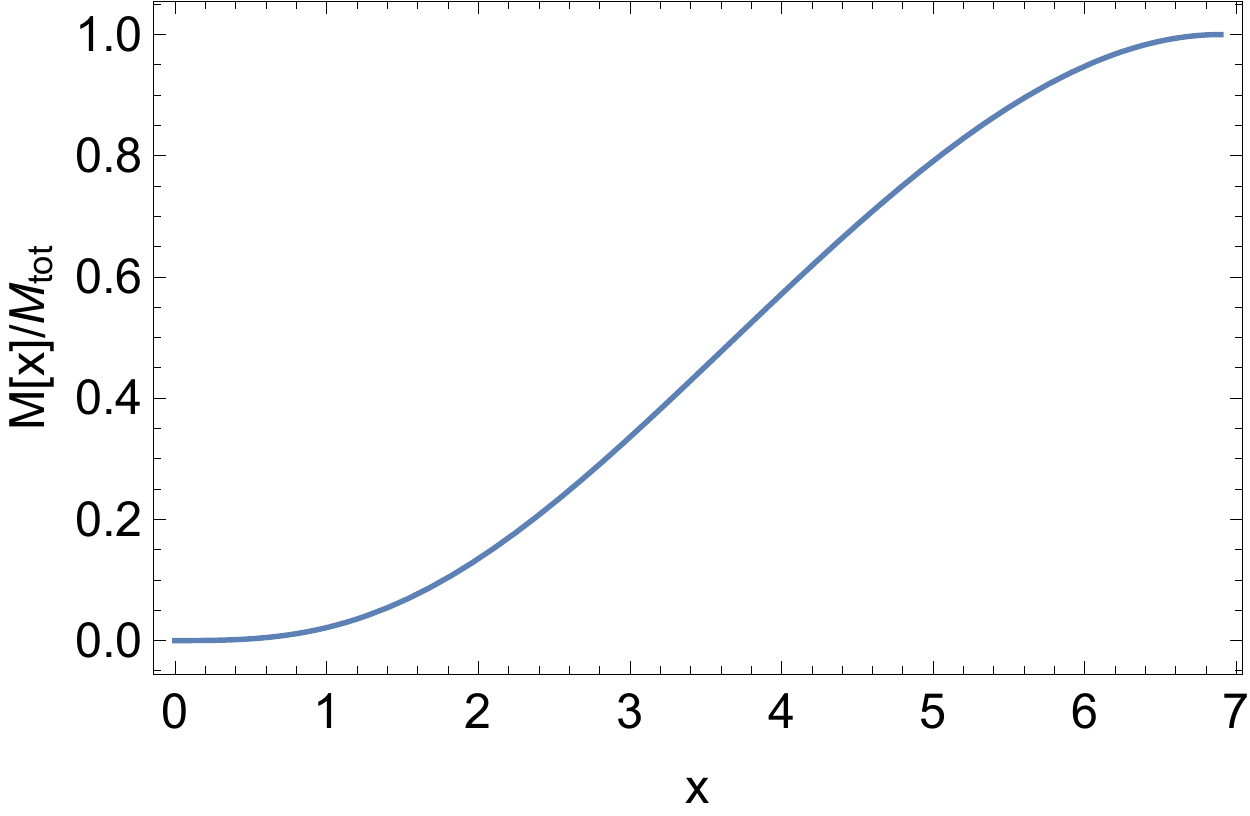}
\includegraphics[width=.32\textwidth]{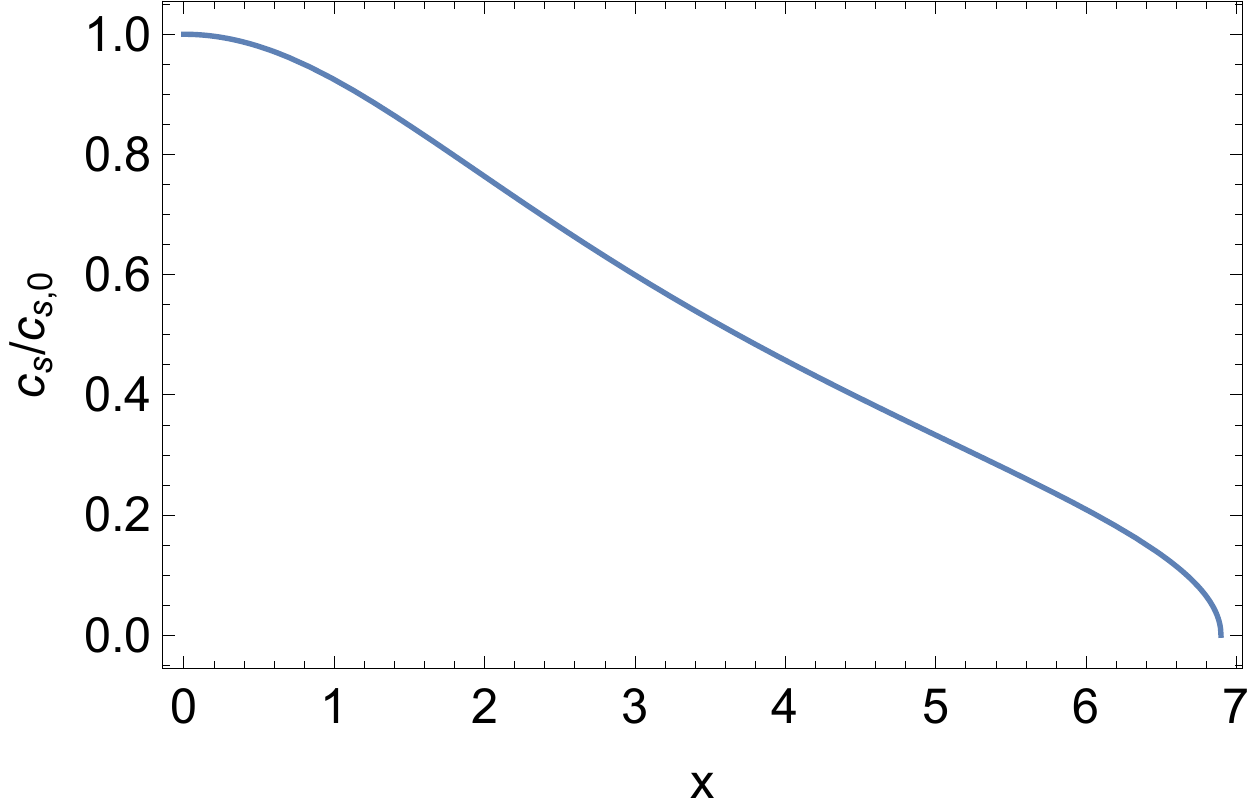}\\
\includegraphics[width=.32\textwidth]{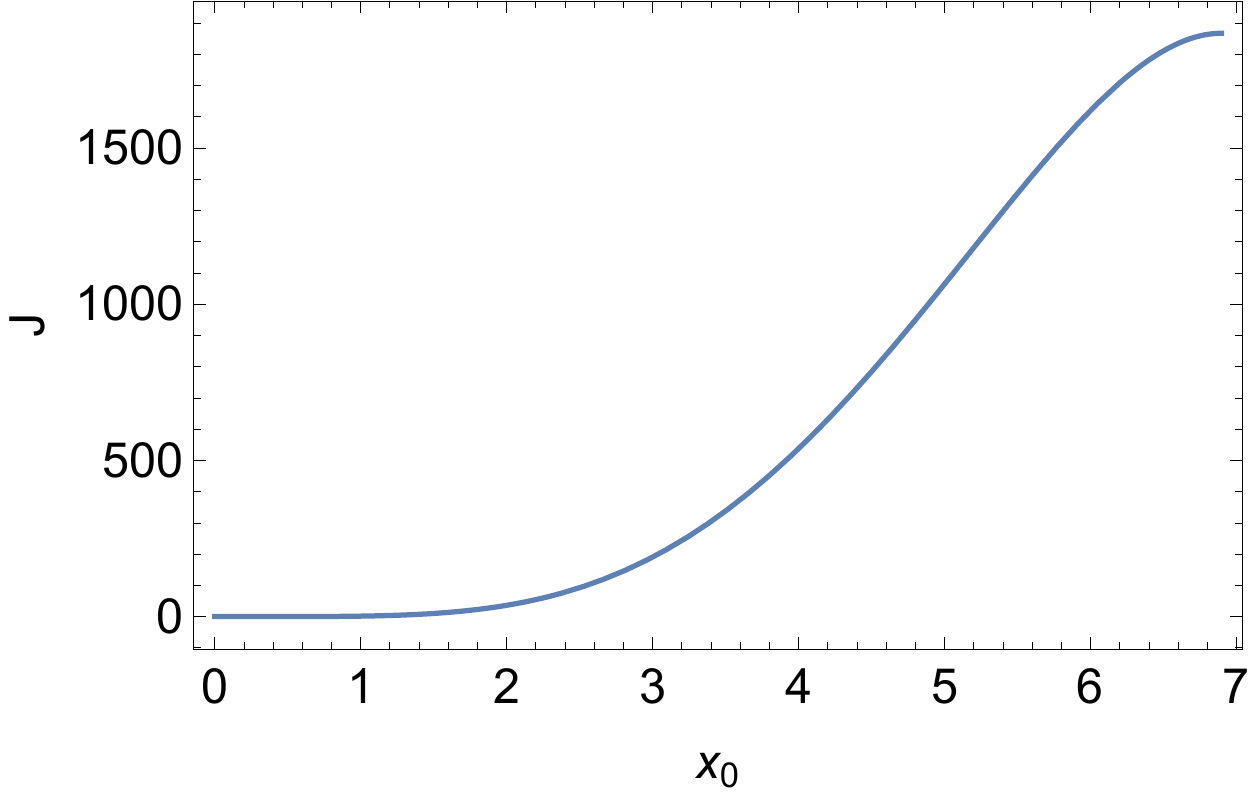}
\includegraphics[width=.32\textwidth]{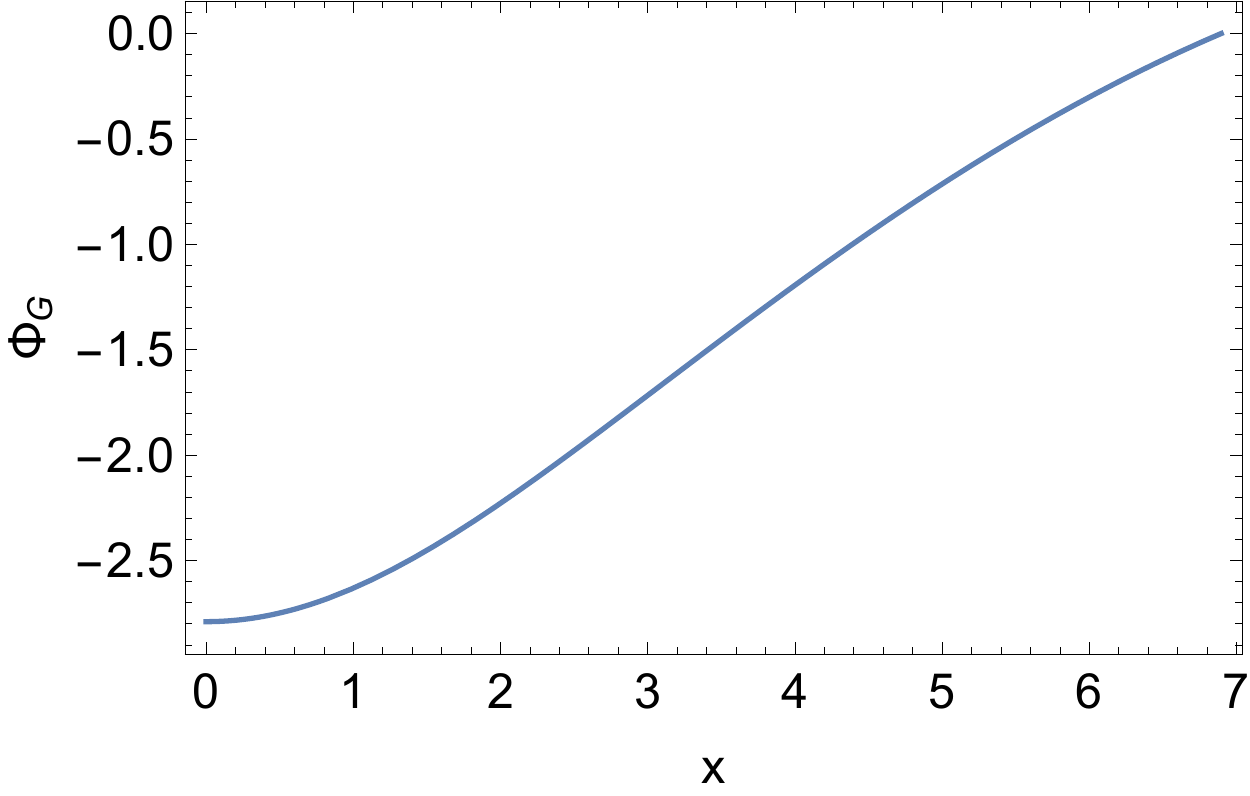}
\includegraphics[width=.32\textwidth]{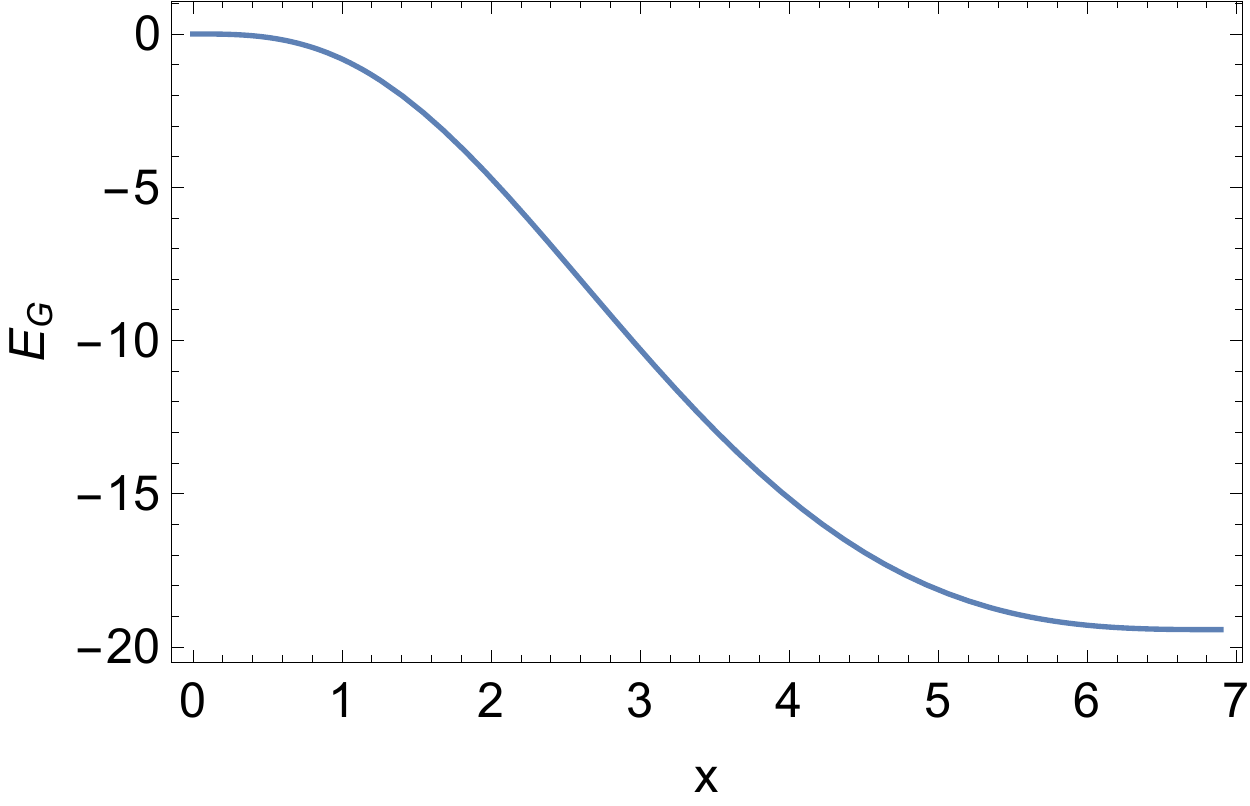}
\caption{Lane-Emden polytropic sphere for $n=3$. Plotted are: Top row:  density (left panel), relative mass (center panel) and sound speed as function of $x$. Bottom row: Left panel -  moment of inertia up to $x_0$; Center panel - gravitational potential as a function of radial coordinate $x$, Right panel -  integrated gravitational energy inside $x$.  }
\label{LaneEmden} 
\end{figure}

For $n=3, \, (\gamma=4/3)$ the surface is located at  $x_{max} = 6.90$ and the total mass is $M_{tot} = 178.3$ (in dimensionless unites).  The   density $\theta$, relative mass and sound speed as function of $x$ are plotted in Fig. \ref{LaneEmden}.
Sound speed is $c_s= c_{s,0} \sqrt{\theta}$.
Moment of  inertia  up to $x_0$ is
\be
J(x_0) = {8 \pi \over 3} \int_0^{x_0} x^4 \theta(x) dx
\ee
(rotational energy unto $x_0$ is $J \Omega_{WD}^2 /2$). Total value $J_{tot} = 1867$. 

To consider the time-evolution we need to normalize the dimensional factors in the Lane-Emden to a particular WD model. Take $M_{WD}= 1.4 M_\odot$ $R_{WD} = 3000 $ km. 
The sound speed at the center is
\be
c_{s,0} = \sqrt{ 4 \pi G \rho_c \over n+1} r_0 =  1.05 \times 10^8 {\sqrt{\frac{M_{WD}}{M_{Ch}}} \over  \sqrt{R_{WD}/3000 \,{\rm km}}} \, {\rm cm\,  s}^{-1}
\ee

\begin{figure}[h!]
\centering
\includegraphics[width=.99\textwidth]{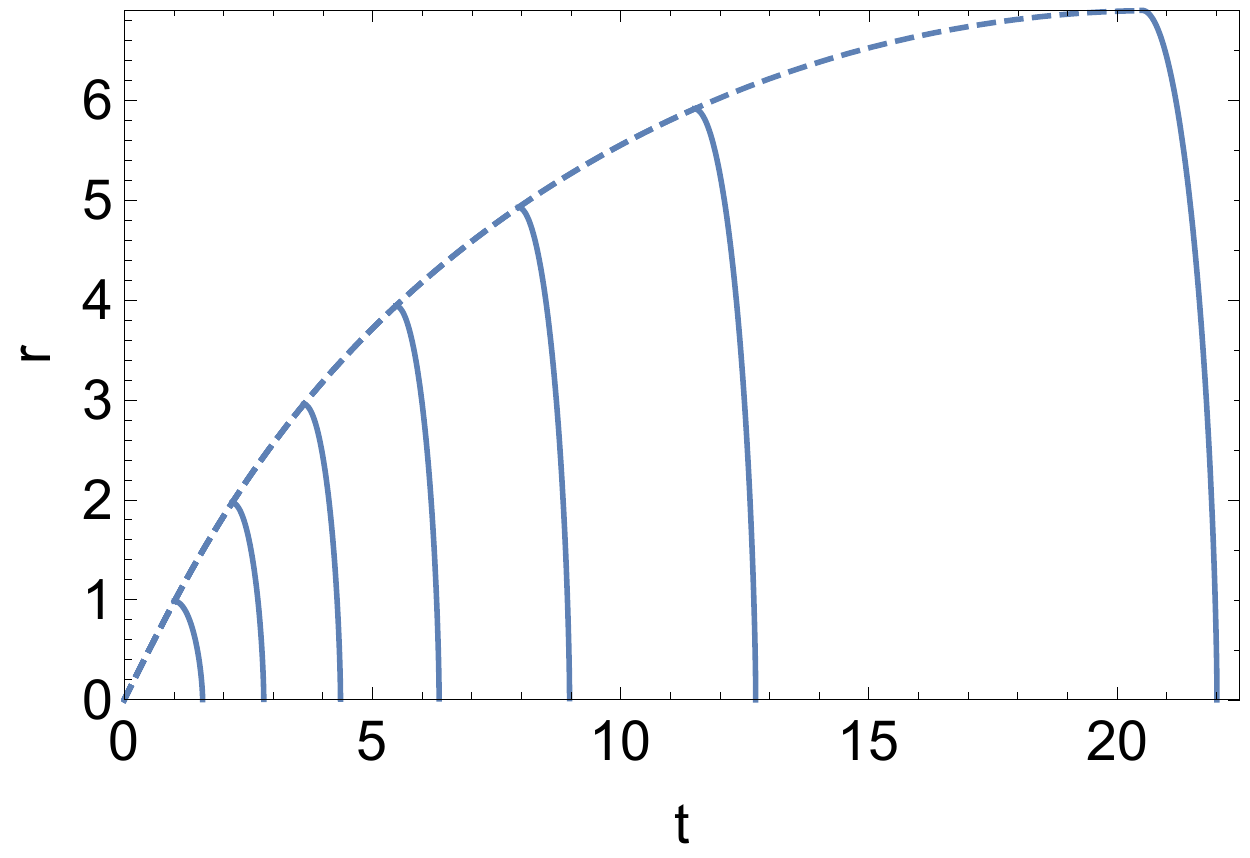}
\caption{Dynamics of the rarefaction wave $r_{RF}(t)$ (dashed line; the surface is reached at $t=20.89$) and  post-rarefaction wave free fall (solid lines are world lines of different elements), dimensionless units, see Eq. (\ref{r0}) and (\ref{tau}). }
\label{RFofx} 
\end{figure}

Rarefaction wave propagates with sound speed. Integrating $\partial_t r_{RF} = c_s(r_{RF})$, we find  the motion of the rarefaction wave, Fig. \ref{RFofx}. After  the  rarefaction wave reaches a point, matter is in a state of near free-fall. Velocity at each point $x$ of the material that started at $x_0$ is then
\be
v= - \sqrt{ 2 M(x_0) (1/x-1/x_0)},
\ee
see  Fig. \ref{RFofx}. 


\subsection{Accretion shock and the bounce}

Let's assume that all the gravitational energy released during collapse is converted into heat and conserved during the evolution (no radiative losses). Knowing the distribution of density $\rho(r)$, parametrized by $\theta(x)$, we can calculate the gravitational potential $\Phi(r)$,
\be
\frac{1}{r^2} \partial_r ( r^2 \partial_r \Phi)  = \rho(r)
\ee
 and gravitational energy up to radius $r$, 
 \be
 E_g  = - \frac{4\pi}{2} \int_0^r  \Phi \rho r^2 dr,
 \ee
 Fig \ref{LaneEmden}.

Let us next estimate the ram pressure created by the infalling material. Assume that the rarefaction waves reaches radius $r$ at times $t$. The mass of the shell located between $r$ and $r+ c_s dt$ is $dm = 4 \pi r^2 \rho(r) c_S dt$. This mass will be accreted during interval 
\be
\Delta t = \frac{\partial t_{ff}}{\partial r} c_s (r) dt
\ee
Thus, the ram pressure created by material falling from radius $r$ is 
\be
p_{ram} = \frac{r^2 v \rho}{r_{in}^2 ({\partial t_{ff}}/{\partial r})}
\ee
For $r_{in} \ll r$, $ v = \sqrt { 2 G M_r/r_{in}}$, where, again, $M_r$ is the mass enclosed within the initial radius $r$. Using (\ref{tff}) we find
\be
p_{ram} =\frac{8 G }{3\pi} \frac{r^{3/2} M_r^{2/3} \rho}{ r_{in}^{5/2} \partial_r (r/M_r^{1/3}))}
\label{pram}
\ee
This is the  ram pressure created at  radius $r_{in} \ll r$  by 
a material located initially at radius $r$, Fig. \ref{pram}.

\begin{figure}[h!]
\centering
\includegraphics[width=.49\textwidth]{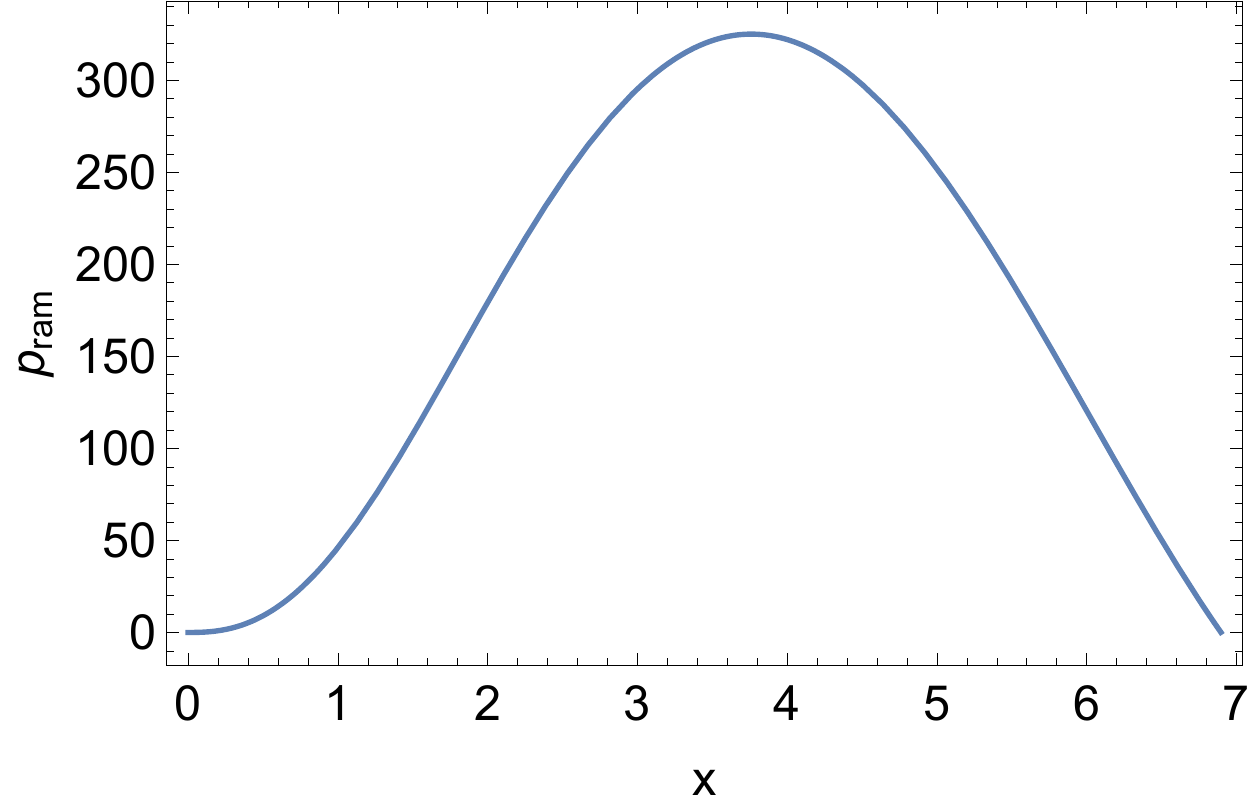}
\includegraphics[width=.49\textwidth]{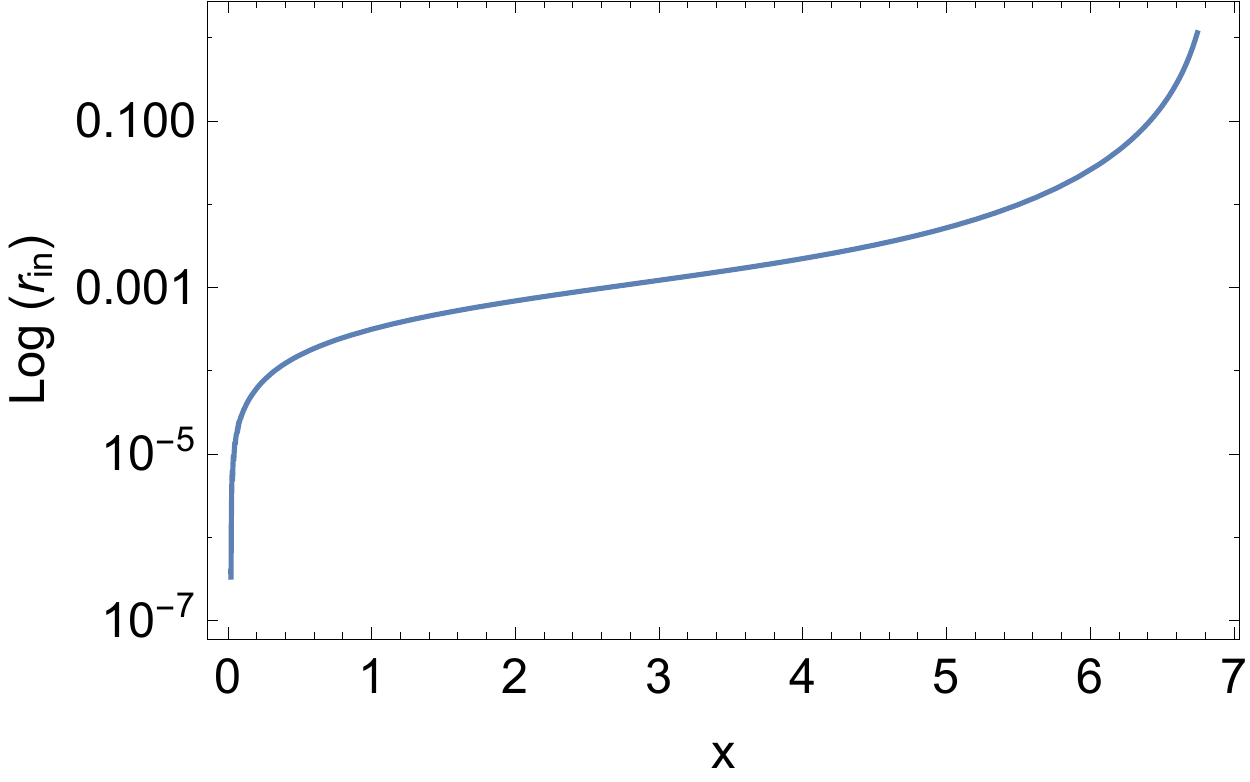}
\caption{ Ram pressure  created by infalling material  initially at  dimensionless radius $x$ (left panel) and the location of the accretion shock (right panel). Large values of $r_{in}$ at the end of the collapse correspond to the shock bounce. }
\label{pram1} 
\end{figure}

To estimate the size of the shocked region, we balance the kinetic energy density brought in by the accreting matter, $ E_g/(4 \pi r_{in}^3 /3)$ with ram pressure (\ref{pram1}). We find
\be
r_{in} =\frac{81}{1024} \frac{E_g \left( \partial_r (r/M_r^{1/3})\right)^2}{G^2 r^3 M_r^{4/3} \rho^2}
\label{rin}
\ee
This expression locates the accretion shock at time when material from radius $r$ is accreted, Fig. \ref{pram1}.

For most of the accretion time the radius of the shock increases approximately linearly with time. But as the surface approaches the shock the ram pressure decreases dramatically.
Since near the surface the density approaches zero linearly (in fact, $\rho \propto  0.044 (x_{max} - x)$), we have $r_{in} \propto (x_{max} - x)^{-2}$ - this is the bounce.

\subsection{Rotation during collapse}

Let's consider rotational evolution during collapse. As a fiducial estimate, let's assume that WD of $R_{WD} =3000$ km collapses to $R_{NS} = 30$ km, by a factor 1/100, to 
$x_{min} = x_{max}/100$. (Further contraction is expected during cooling.)

We employ the following model of the collapse: (i) we separate the star into weakly rotating outer parts, unaffected by the rarefaction wave, (ii) a free-fall region, (ii) a core. We assume that the core has a radius much smaller than the initial configuration, so that at each moment its mass and angular speed are determined by the amount of the material and angular momentum that has reached the point $r=0$ before a given moment. We assume that the radial infall dynamics is unaffected by the rotation (hence the requirement of sufficiently slow rotation discussed above.)

Let's assume that the central object has the same gyro-ratio  as the initial WD: $\eta_{0} =0.22$. Then, after material from $x_0$ accreted on the central object, conservation of angular momentum gives
\be
J(x_0) \Omega_{WD} = \eta_0  x_{min}^2 M(x_0) \omega_c
\ee
\begin{figure}[h!]
\centering
\includegraphics[width=.49\textwidth]{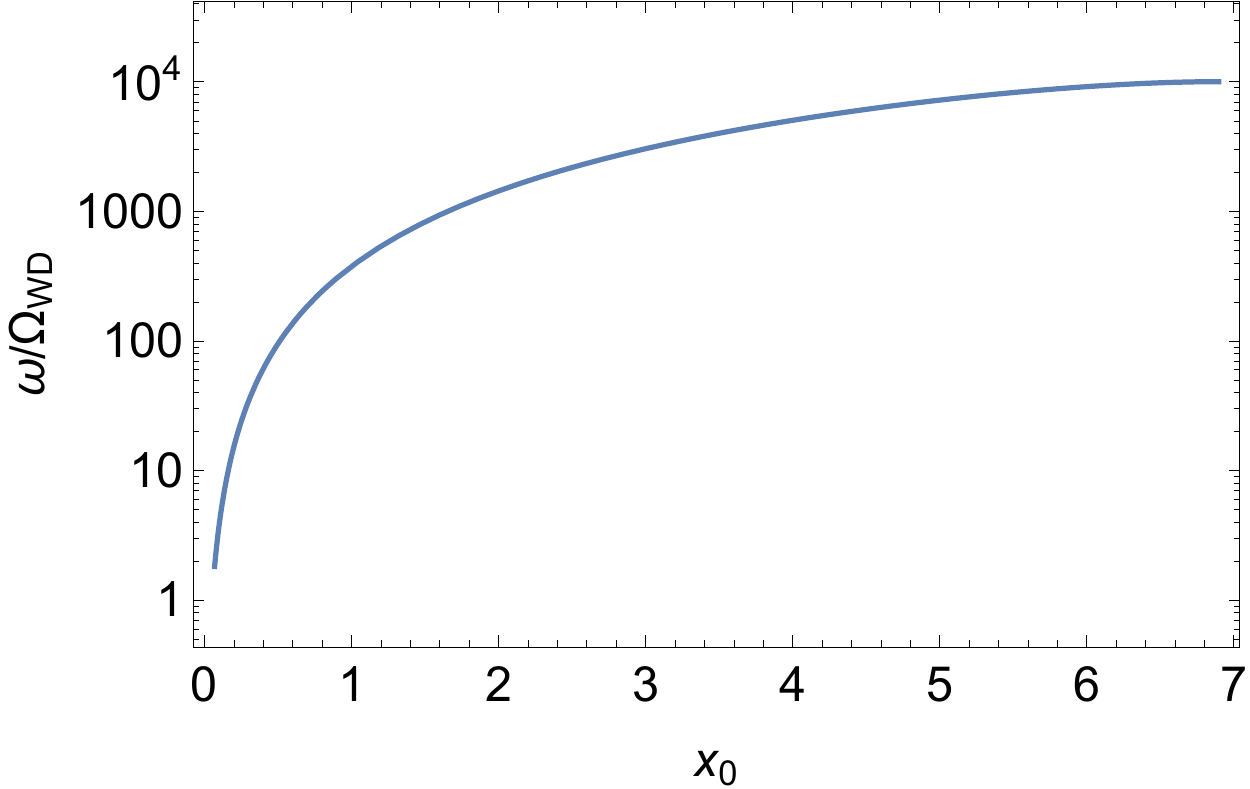}
\includegraphics[width=.49\textwidth]{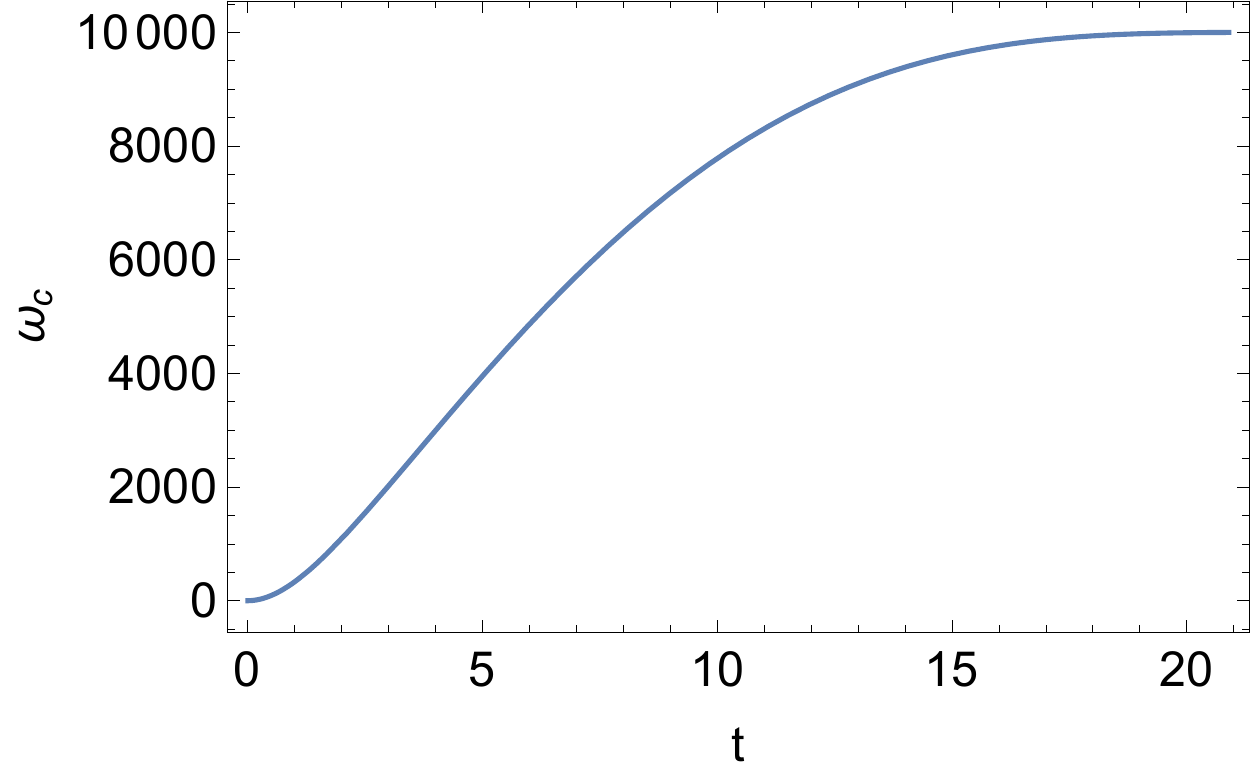}
\caption{Left Panel: Angular velocity of the central object as function of the initial position of the accreted  shells. Right panel: angular velocity of the central object as function of  time. Maximal value is $\om/\Omega_{WD}= (x_{max}/x_{min})^2 =10^4$.}
\label{omegaofx0} 
\end{figure}

Let us then calculate the temporal evolution of the angular velocity of the central object. It may be demonstrated that (in physical units) the free-fall time $t_{ff}$ from a location $r$ that encloses mass $M_r$ is
\be
t_{ff}= {\pi \over 2 \sqrt{2}} {r^{3/2} \over \sqrt{ G M_r}}
\label{tff}
\ee
Interestingly, for a polytropic EoS this gives a final time even for $r=0$ (since  $M_r \propto r^3$ for small $r$, $t_{ff} = 0.96 \tau$. To account for this mathematical oddity, we subtract from (\ref{tff}) a value for $r=0$. In terms of the dimensionless time,
\be
t_{ff}\equiv {t_{ff}\over \tau} = 1.968 {x_0^{3/2}\over M(x_0)} - 0.96, 
\ee
For the surface layer $t_{ff}(x_{max}) = 1.70$. 
The angular velocity of the core is then, Fig. \ref{omegaofx0}, 
\be
\om_c(t) = {J(t) \over \eta_0 x_{min}^2 M(t)}.
\label{omc}
\ee

\subsection{Magnetic field amplification}

Previously \Bf\ amplification during collapse was considered by \cite{1995ASPC...72..301T}. They argued that {\it if} the accretion is direct, without the formation of the accretion disk, \Bf\ can be amplified by an $\alpha-\omega$ dynamo, while the star can be spun to millisecond periods. A simple flux conservation gives
$B_{NS} \approx (R_{WD}/R_{NS})^2 B_{WD} \approx 10^{12} {\rm G} B_{WD, 8}$. \cite{1995ASPC...72..301T} argued that twisting of the \Bf\ during collapse and ensuing dynamo action can bring the \Bf\ to magnetar values.
In what follows we use the above-given calculations of the infall, add rotation, and estimate the resulting \Bf. We find that even without dynamo action the \Bf\ can be twisted to magnetar values.

As the star collapses, differential rotation will lead to amplification of the toroidal \Bf.
To estimate the  \Bf\ amplification, we assume that the initial configuration is mostly poloidal. Then the resulting toroidal field will be larger than the  final poloidal field approximately by the difference in the number of turns between  the core and the outer layer. 
To estimate the the number of turns that the core makes, we integrate $\om_c(t)$, Eq. (\ref{omc}) over collapse time,
\be
N_c= {1\over 2 \pi} \int \om_c(t) dt = 2.2 \times 10^4
\label{Nc}
\ee

Before the RF reached the outer layer, it rotates with the initial $\Omega_{WD}$; in addition, as the outer layer falls onto the star it's rotational velocity increases. We find for the number of turns of the outer layer
\be
N_{outer}= 6.7
\ee
The fact that the total winding of the toroidal \Bf, Eq. (\ref{Nc}), reaches $\sim 10^4$ has a simple order-of-magnitude explanation: the collapse takes about 2 dynamical times from the center to the surface (for RF and the ensuing free-fall). Typically the core rotates with $\om_c \sim (x_{out}/x_{in})^2 \Omega$, while the outer layer rotates with $ \Omega$. For  $x_{out}/x_{in} \sim 100$ this gives  $\om_c \sim 10^4$.

Thus, the toroidal  \Bf\  can be $\sim 10^4 $ times higher than the poloidal. In addition, poloidal \Bf\ will be amplified by flux conservation. 
For example, if we start with $B_{WD} \sim 10^6$ G, flux conservation will give a factor $(x_{\max}/x_{min})^2 \approx 10^4$, while differential rotation will further boost that by 
$\sim  2.2 \times 10^4$, reaching magnetar-like values of $B \geq B_Q$.

\section{The extended emission (EE) and the final NS}
\label{EE}

\subsection{Shell forms a Keplerian disk around newly formed NS}

As our basic scenario of the post-merger evolution we chose the left column of  Fig. \ref{outline-WD-2} - no prompt collapse, CO shell burning  and AIC after the core exceeds $M_{Ch}$.  For definiteness let's assume that the mass of the shell at the moment of AIC is $M_d\sim 0.5 M_\odot$, its typical radius is $R_s \sim 5 \times 10^9$ cm \citep[see \eg][]{2012ApJ...748...35S}, and  it is rotating with the period $P_s \sim 10^3$ sec. The \NS\ has its \Bf\ amplified to nearly critical values and is spinning at few milliseconds.

The material of the shell has a proper angular momentum $\sim \Omega_s R_s^2$. After the loss of support from the core, the shell will form a disk at the radius $R_K$ where this  proper angular momentum corresponds to Keplerian rotation, 
\ba && 
\Omega_s R_s^2 = \sqrt{G M_{NS} R_K}  \rightarrow 
\nn &&
R_K = \frac{R_s^4 \Omega_s^2}{ G M_{NS}}= 10^8\, \left(\frac{R_s}{5 \times 10^9 \, {\rm cm}} \right)^4 \,  \left( \frac{P_s}{10^3 \, {\rm sec}}\right) ^{-2} \,{\rm cm}
\nn &&
\Omega_K= \frac{(G M_{NS})^2}{R_s^6 \Omega_s^3}= 10\, \left(\frac{R_s}{5 \times 10^9 \, {\rm cm}} \right)^{-6} \,  \left( \frac{P_s}{10^3 \, {\rm sec}}\right) ^{-3} \,{\rm rad \, s^{-1}}
\ea
Thus, the newly formed disk rotates with a period $P_K$ of the order of a second.

It is expected that the accretion time, \eg\ within the $\alpha$-prescription, 
\be
\tau_{acc} \sim \frac{P_K}{\alpha} \left( \frac{R_d}{H_d}\right)^2
\ee
will be $\sim 1/\alpha\approx  10-100$ times longer than the  period (assuming that $R_d \sim H_d$ and $\alpha =10^{-2} -10^{-1} $).
 Thus the accretion rate will be of the order
\be
\dot{M} \approx \alpha \frac{ \Delta M}{P_{K}} \approx 10^{-3} M_\odot {\rm s} ^{-1}
\label{dotM}
\ee
Typical accretion time scale
\be
\tau \sim \frac{M_d}{\dot{M}}\approx \frac{P_{K}}{\alpha} \approx  100  {\rm sec}
\ee
This is the extended emission (EE).

Accretion of material onto the NS is a famously complicated problem. As basic estimate we note, that the accretion rate (\ref{dotM}) is so high, that the \Alfven radius could be smaller than the NS radius
\be
\frac{r_A}{r_{NS}}= 
\frac{2^{4/7} \pi ^{2/7} B_{{NS}}^{4/7} R_{{NS}}^{5/7}}{{G M }^{1/7} 
   {\dot{M}}^{2/7}} = 0.9 \left(\frac{B_{NS,}}{B_Q} \right)^{4/7}
   \ee
Thus,  direct accretion on the NS can occur even for magnetar-like fields ($B_{Q}$ is  critical quantum field \Bf). 

For somewhat different sets of parameters, the shell will be ejected in a propeller regime \citep{1992ans..book.....L,2006ApJ...646..304U}, when the magnetospheric radius is larger than the corotation radius and smaller than the light cylinder radius. In this case most of the matter is expelled radially in the equatorial plane by the rotating magnetosphere of the star.

 The expected upper limit on the EE  luminosity, accretion-powered,  can be estimated as
\be
L_{EE} \approx \eta_a \dot{M}  c^2 \frac{R_{{NS}}}{R_{G}}= 7 \times 10^{49}\,\eta _{-1} \,  {\rm erg s}^{-1}
\ee
where $\eta _{a,-1}= \eta_a/10^{-1}$ is the efficiency of converting accretion power into radiation. This is comparable to  the observer EE power.

We also point out that  accretion  in this case happens in an interesting, not yet explored regime: very high $\dot{M}$ on to rapidly spinning ultra-magnetized \NS. In certain regimes accretion will proceed in the propeller regime, so that a large amount of rotational energy of the \NS\ is given to the ejected material  \citep{2011ApJ...736..108P}.
 
 \subsection{Afterglows: pulsar-like  termination shock in fast wind}
 
 As a result of the AIC a highly spinning, with a period few  milliseconds, NS is born. At the same time \Bf\ can be amplified to magnetar fields. 
 Importantly,  after all the secondary material is accreted the nature of the collimation changes: isolate \NSs\ lose most of it's rotational energy {\it in the equatorial plane},
 with luminosity $L\propto \sin^2 \theta$, where $\theta $ is the polar angle. The highly magnetized relativistic wind produced by a central \NS\ will interact with the fairly dense newly ejected material and dense pre-AIC wind,  producing  $X$-ray afterglow in the  {\it highly magnetized  reverse shock}, in a way similar to the case of afterglows from long GRBs  \citep{2017ApJ...835..206L}.

\section{Discussion}

We advance a model of short GRBs originating from unstable mass transfer (merger of) from a CO WD onto a heavy ONeMg WD. The disrupted  CO WD enters a shell burning stage, adds material to the core of primary, which experiences  the accretion induced collapse. During AIC the  \Bf\ is amplified, while the remaining shell provides a collimation of the outflow. Accretion continues onto the newly formed NS from the disrupted companion, producing extended emission. At later stages pulsar winds interact with the pre-collapse wind (and possibly with the  shell ejected in a propeller regime of accretion) and produces afterglows by particle acceleration at the termination shock, similar to Pulsar Wind Nebulae and what has been suggested by long GRBs \citep{2017ApJ...835..206L}.

The model  explains both the prompt GRB stage - around the bounce time  - and  the  extended emission - due to shell accretion. Additional flaring activity may be produced by the newly born \NS\ via magnetar-like flares. In addition  mild optical signal is expected during the AIC of the WD (\eg\ material ejected during the bounce interacting with the remaining accretion disk). We do not expect in this case the GRBs to be associated with a strong gravitational wave signal.

Qualitatively, the model allows for a larger variety of properties of short GRBs than the standard NS-NS merge paradigm. The most important  macroscopic variable parameters are the masses of the WDs before the merger (this depends on the initial masses and separation), and pre-AIC rotational periods of the shell and the core. In addition, there is a possibility that AIC occurs during the active stage of the CO WD disruption.

Importantly, it is  {\it required} that the proposed scenario be very rare, accounting to at most only few percent of the WD mergers (or at most 10\% of super-\Ch\ mergers). 
Other (majority of) channels may lead either to SN Ia explosions or very weak transients. Only the right combination(s) of initial masses and separations should lead to short GRBs associated with WD mergers.

 The present model postulates a fairly dense  surrounding  of short GRBs (due to the presence of unburnt shell and the powerful pre-AIC wind)  in contrast to a very clean circumburst environment of NS-NS mergers. We view it as a strong point of the model - both long and short GRBs have tenths to few solar masses of the material in the immediate surrounding of the explosion (smaller for shorter bursts). This leads to similarly looking early  afterglows, just somewhat less energetic in case of short GRBs. The model also qualitatively explains these long-short GRBs similarities as both being powered by an ultra-relativistic highly magnetized wind, produce, \eg  by a millisecond highly magnetized \NS\ in both cases. 

\begin{figure}[h!]
\centering
\includegraphics[width=.99\textwidth]{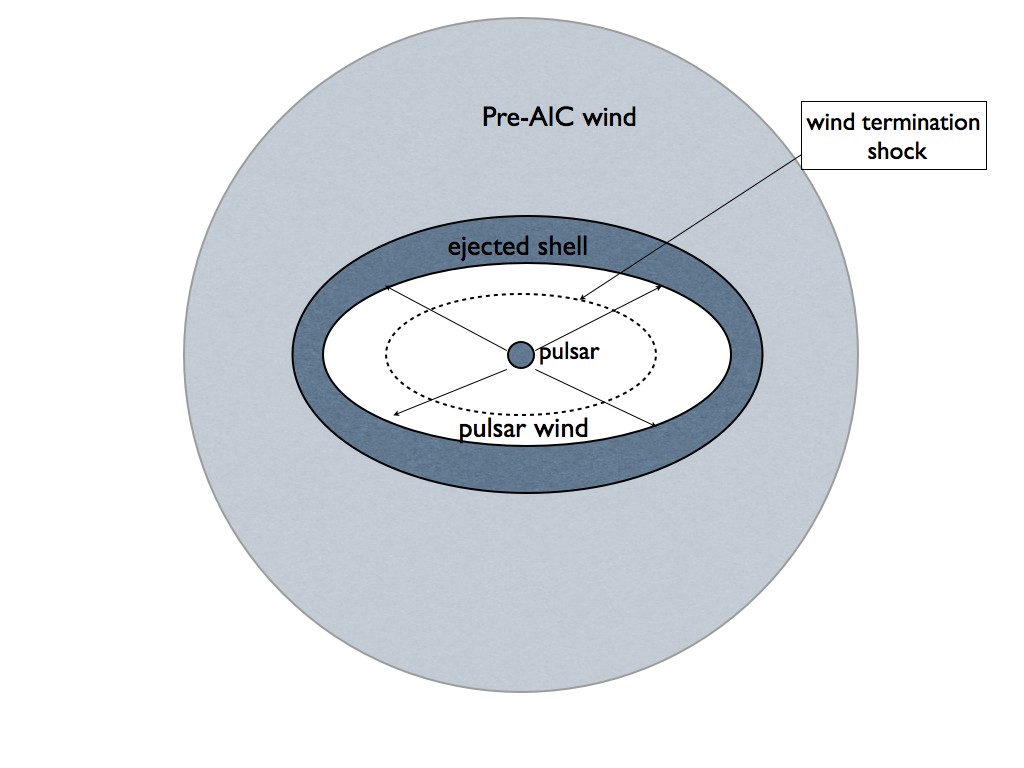}
\caption{Immediate surrounding of a short GRB: the remaining NS generates pulsar-like wind (highly relativistic, highly magnetized), that interacts with the shell ejected in a propeller regime, and the wind from the shell burning stage.  Both pulsar winds and the ejected shell are equatorially collimated. Early afterglows are produced at the wind termination shock, similar to the case of long GRBs discussed by \protect\cite{2017ApJ...835..206L} }
\label{SurrondingSGRB} 
\end{figure}

The present model offers a number of possibly  interesting developments: 
\begin{itemize}
\item Interaction of the AIC bounce ejecta with the envelope. We expect that $ M_{ej} \sim 10^{-3} -10^{-2} M_\odot $ are ejected during AIC with mildly relativistic velocities, \eg\ with $v_{ej} \sim c/3$, the sound speed in relativistic fluid. The total energy of the ejecta can be $\sim 10^{50} - 10^{51}$ ergs. This ejecta immediately runs into $M_s \sim $ few  $\times 10^{-1} M_\odot $ shell. By  conservation of the momentum, the resulting shell will be moving with velocity 
\be
v_s \sim \frac{c}{3} \frac{M_{ej}}{M_s} \sim 10^9 \, {\rm cm \, s}^{-1}
\label{vs}
\ee
The shell will be heated to 
\be
T \sim \frac{m_p c^2}{18} \frac{M_{ej}}{M_s} \sim {\rm few\,  MeV}
\ee
Thus, the interaction of the AIC bounce ejecta with the shell will produce 
a very dirty GRB, with long time scales and lower photon energies. 
\item Shock breakout from the shell. The  AIC bounce ejecta will launch a shock in the shell. During the breakÐout of the shock, the top material will be accelerated to mildly relativistic velocities, even though the bulk of the shell material moves non-relativistically, Eq (\ref{vs})
\item Interaction of the shell with the pre-AIC wind  produces  directionally anisotropic radiation features, Fig. \ref{SurrondingSGRB}. For example, shell ejection is a  propeller regime and pulsar-generated wind will be equatorially collimated, yet in ``strong propeller'' regime \citep{2006ApJ...646..304U}  collimated magnetically dominated outflow can be present as well.
\end{itemize}

We would like to thank Chris Fryer, Natasha Ivanona, Mansi Kasliwal, Marten van Kerkwijk, Patrick Motl, Kelly Lepo,  Lorne Nelson, Thomas Tauris,  Sterl Phinney,
Tony Piro, Philipp Podsiadlowski,  Jan Staff, Ken Shen, and organizers of  NORDITA workshop The Physics of Extreme-Gravity Stars, and Samaya Nissanke in particular. 
This work was supported by   NSF  grant AST-1306672 and DoE grant DE-SC0016369.

\bibliographystyle{apj} 
  \bibliography{/Users/maxim/Home/Research/BibTex}   
  
  \end{document}